\documentclass[aps,11pt,amssymb,prd,preprintnumbers,amsmath,amsfonts,nofootinbib]{revtex4-1}
\usepackage{latexsym}
\usepackage[latin1]{inputenc}
\usepackage{amsmath}
\newcommand{\ba}{\begin{eqnarray}}
\newcommand{\ea}{\end{eqnarray}}

\newcommand{\be}{\begin{equation}}
\newcommand{\ee}{\end{equation}}
\newcommand{\pa}{\partial}

\newcommand{\nn}{\nonumber}

\newcommand{\Tr}{{\rm Tr }}

\newcommand{\vts}{{\slashed{\tilde{v}}}}

\newcommand{\WT}{\xrightarrow{\ \ \textrm{WT} \ \ } }
\usepackage{hyperref}  % for faster clicking around equations

\usepackage{dcolumn}% Align table columns on decimal point
\usepackage[dvips]{graphicx}
\usepackage{color}
\usepackage{slashed}
\usepackage{ulem}
\definecolor{stcol}{rgb}{1,0,1}

\begin{document}
                                                                                
\date{\today}

\title{Chiral kinetic theory with small mass corrections and quantum coherent states}
\author{Cristina Manuel}
\email{cmanuel@ice.csic.es}
\affiliation{Instituto de Ciencias del Espacio (ICE, CSIC) \\
C. Can Magrans s.n., 08193 Cerdanyola del Vall\`es, Catalonia, Spain
and \\
 Institut d'Estudis Espacials de Catalunya (IEEC) \\
 C. Gran Capit\`a 2-4, Ed. Nexus, 08034 Barcelona, Spain}
 
\author{Juan M. Torres-Rincon}
\email{torres-rincon@itp.uni-frankfurt.de}
\affiliation{Institut f\"ur Theoretische Physik, Johann Wolfgang Goethe-Universit\"at, Max-von-Laue-Strasse 1, D-60438 Frankfurt am Main, Germany}
\begin{abstract}
We study the effect of a small fermion mass in the formulation of the 
on-shell effective field theory (OSEFT). This is our starting point to derive 
small mass corrections to the chiral kinetic theory. In the massless case, 
only four Wigner functions are needed to describe positive and negative 
energy fermions of left and right chirality, corresponding to the vectorial components of a fermionic two-point 
Green's function. As soon as mass correction are introduced, tensorial 
components are also needed, while the scalar components strictly vanish 
in the OSEFT. The tensorial components are conveniently parametrized in 
the so-called spin coherence function, which describe quantum coherent 
mixtures of left-right and right-left chiral fermions, of either 
positive or negative energy. We show that, up to second order in the energy expansion, vectorial and tensorial components are decoupled, and obey the same dispersion law and transport equation, depending on their respective chirality.
We study the mass modifications of the reparametrization invariance of 
the OSEFT, and check that vector and tensorial components are related by the associated symmetry transformations. We study how the macroscopic properties of the system are described in terms of the whole set of Wigner functions, and check that our framework allows to account for the mass modifications to the chiral anomaly equation.
\end{abstract}

\maketitle

\section{Introduction}

In this manuscript we study how small mass corrections affect the chiral transport theory derived from the on-shell effective field theory (OSEFT), following our previous studies in the subject~\cite{Manuel:2014dza,Carignano:2018gqt,Carignano:2019zsh}.

Chiral kinetic theory (CKT) is a well defined framework to study quantum transport phenomena in chiral systems. Initially developed in Refs.~\cite{Son:2012wh,Son:2012zy,Stephanov:2012ki,Chen:2012ca}, alternative derivations have been worked out~\cite{Hidaka:2016yjf,Gorbar:2016ygi,Mueller:2017lzw,Mueller:2017arw,Gao:2018wmr,Hidaka:2018ekt,Gorbar:2017awz,Huang:2018wdl}. However, in most real physical systems fermions are not strictly chiral, but have a small mass. It is thus interesting to consider how small mass effects modify the CKT. Several works have appeared in the literature that consider these mass modifications~\cite{Zhuang:1995pd,Hattori:2019ahi,Gao:2019znl,Weickgenannt:2019dks,Wang:2019moi,Li:2019qkf,Gao:2019zhk,Weickgenannt:2020sit,Yang:2020hri,Liu:2020flb}.

We start our work by studying how a small mass modifies the OSEFT, which was initially developed to describe massless chiral fermions. This is similar to other studies of different effective field theories for massless fermions, such as HDET~\cite{Hong:1998tn} or SCET~\cite{Bauer:2000ew,Bauer:2000yr}, see Refs.~\cite{Schafer:2001za,hep-ph/0505030,Leibovich:2003jd}. As there are several parallelisms between OSEFT and SCET that were already discussed in Ref.~\cite{Carignano:2018gqt}, we find again the same sort of mass modifications here. Remarkably, these affect the so-called reparametrization invariance (RI) of the theory, needed to check that the whole formalism
respects Lorentz symmetry. Understanding the symmetries of the OSEFT will turn out to be essential to give a complete and coherent transport approach.

In Ref.~\cite{Carignano:2018gqt} we derived transport equations for the distribution functions of right and left chiral massless fermions, by focusing on the vectorial components of the OSEFT two-point Green's function, after performing a Wigner transformation and a gradient expansion.  A mass term breaks the chiral symmetry in the original Lagrangian, but in the OSEFT Lagrangian, expanded in powers of the inverse of the energy, this is an effect that only occurs at second order. Scalar components of the two-point Green's function strictly vanish in the OSEFT, while the tensorial components are also needed. The last can be conveniently parametrized in terms of the so-called spin coherence functions, that were first introduced in studies of neutrino transport equations~\cite{Vlasenko:2013fja,arXiv:1605.09383}.
The associated distribution functions describe quantum coherent mixtures of left/right and right/left-handed fermions.  In the massless case, vectorial and tensorial components are completely decoupled. We find that, up to the second order in the energy expansion considered in this manuscript,
the vectorial and tensorial components are also decoupled. We then derive their corresponding collisionless transport equations, which turn out to be identical.

We study the RI of the OSEFT. One of the RI transformations, which is a particular combination of infinitesimal Lorentz boosts and rotations, does not respect the chirality of the OSEFT fields. As a consequence, in the transport framework, the Wigner function associated to vectorial and tensorial components are related after carrying out these combined transformations. In particular, this means that both components are  needed to respect the Lorentz symmetry in this slightly modified CKT.

We  work explicitly how several of the macroscopic quantities are described in terms of both the vectorial and tensorial components, and check that the chiral quantum anomaly, with small mass corrections, is properly described in our framework.

This manuscript is structured as follows. In Sec.~\ref{mOSEFT} we derive the small mass corrections to the OSEFT Lagrangian, noticing that operators that break chirality only appear at second order in the energy expansion. In Sec.~\ref{WigFun} we show the possible Dirac structures  of the OSEFT two-point functions, and thus of the Wigner functions. While there are no scalar structures, as particles and antiparticles are totally decoupled in the OSEFT, vectorial and tensorial are needed. We parametrize the latter in terms of complex functions. In Sec.~\ref{Disp-sec} we derive the dispersion relations of both vectorial and tensorial components, stressing that at the second order in the energy expansion, these are completely decoupled. We derive in Sec.~\ref{Derivation} the transport equations of both components. In Sec.~\ref{RiOSEFT} we discuss how  the RI of the OSEFT is modified with mass corrections, and deduce the corresponding transformations rules of the Wigner's functions.
In Sec.~\ref{macro} we write several of the macroscopic quantities in the system in terms of the Wigner functions. In particular,  in Sec.~\ref{app:scalar} we consider the scalar and pseudo-scalar densities, while in Sec.~\ref{macro-vector} we write the vector and axial-vector currents. In Sec.~\ref{dirac-rep} we make the connection to previous results based on the Dirac equation, by matching the OSEFT spin coherence functions to the corresponding vector functions defined in those works. We then check in Sec.~\ref{chiral-anomaly} that we can reproduce the mass corrections proportional to the pseudoscalar density of the chiral anomaly.  We conclude in Sec.~\ref{Conclu}, stressing that the mass corrections discussed in this manuscript may have an impact on the short time evolution of almost chiral plasmas. We have included in appendices several details of our computations. In App.~\ref{app:Wigner} we recall how the Wigner transformation and gradient expansion are carried out.
In App.~\ref{app:traces} we collect useful values of traces of the two-point function projected over different combinations of Dirac matrices. In App.~\ref{app:details} we give intermediate details needed in the computations of both the dispersion relations and transport equations. In App.~\ref{app:scalarden} we present in full detail the computation
of the scalar density, and show in App.~\ref{app:typeIISigma} how it behaves under one of the RI transformations, showing then that it is a Lorentz scalar. We use natural units in this work, $\hbar=c  =k_B =1$.

\section{A small mass term in OSEFT}
\label{mOSEFT}

Let us check how a small mass affects the OSEFT as originally formulated in~\cite{Manuel:2014dza,Manuel:2016wqs}. The starting Lagrangian describing a charged fermion is
\be
{\cal L}_{v} = \bar{\psi}_v \left( i\slashed{D} -m \right) \psi_v  \ ,
\ee 
where $D_\mu = \pa_\mu + i e A_\mu (X)$ is the covariant derivative. We assume that the mass $m$ is such that $m \ll E$, so that the it can be treated as a perturbative parameter. We review here the basic ingredients of the OSEFT, and the corrections introduced by a small mass term. Let us stress that
 SCET has also been studied in the presence of a small quark mass \cite{hep-ph/0505030,Leibovich:2003jd}. We will find the same sort of mass modifications here, also in the RI transformations, which will be discussed later (see Sec.~\ref{RiOSEFT}). 

We split the fermion momenta into  on-shell and  residual parts,
\be
q^\mu = p^\mu + k^\mu= E v^\mu + k^\mu \ ,\label{eq:Qpart}
\ee
where $E = p \cdot u$, is the energy in the frame associated to the time-like vector $u^\mu$, $u^2 =1$.
We  consider as $v^\mu$ is a light-like vector, and define another light-like vector $ {\tilde v}^\mu$, such that
\be u^\mu = \frac{v^\mu + {\tilde v}^\mu}{2} \ . \label{eq:uvec}\ee
We thus assume
\be
\label{light-condition}
v^2 = {\tilde v}^2 = 0 \ , \qquad  v \cdot {\tilde v} =2 \ .
\ee
Thus $u \cdot v =1$ and $u^2 =1$ are automatically fulfilled. In the local rest frame (LRF) we take $u^\mu = (1,0,0,0)$.

The Dirac field  can be written as
\begin{equation}
\psi_{ v, \tilde v}=e^{-iEv\cdot x}\left( P_v \chi_{v}(x) +  P_{\tilde v} H_{\tilde v}^{(1)}(x)\right)+
e^{iE\tilde{v}\cdot x} \left( P_{\tilde v} \xi_{\tilde v}(x)+P_v   H_{v}^{(2)}(x) \right)\ ,
\label{eq:Fields}
\end{equation}
where  the basic OSEFT quantum fields  obey
\begin{eqnarray}
&P_v \chi_v = \chi_v &\ , \qquad P_{\tilde v} \chi_v = 0 \ , \\
&P_{\tilde v} \xi_{\tilde v} = \xi_{\tilde v} & \ , \qquad P_v \xi_{\tilde v} = 0 \ ,
\end{eqnarray}
and the particle/antiparticle projectors are expressed as
\begin{eqnarray}
P_v & = & \frac{1}{2}  \slashed{v} \, \slashed{u} =  \frac{1}{4}  \slashed{v} \,  \slashed{\tilde v} 
\ , \label{eq:AProj-bis1}\\
P_{\tilde v} & = & \frac{1}{2}  \slashed{\tilde v} \, \slashed{u}  = \frac{1}{4}   \slashed{\tilde v} \,\slashed{v}     \ ,
\label{eq:AProj-bis2}
\end{eqnarray}
where we used that  $\slashed{v} \,\slashed{v} =  \slashed{\tilde v} \, \slashed{\tilde v} = 0$. Note that
$P_v ^\dagger = \gamma_0 P_{\tilde v} \gamma_0$.

It is possible to integrate out the $H_{\tilde v}^{(1)},H_v^{(2)}$ fields of the QED Lagrangian~\cite{Manuel:2014dza}, to have an effective theory for the fields $\chi_v$ and $\xi_{\tilde v}$ only. We will focus on the particles from now on, as the Lagrangian for antiparticles can be deduced by simply exchanging $E \rightarrow -E$, and $v \leftrightarrow{\tilde v}$~\cite{Manuel:2014dza}.

In an arbitrary frame, and focusing on the particle sector,
\begin{align}
\nonumber
{\cal L}_E & =  \sum_{\bf v} \left[ \bar \chi_{v} (x)  i v \cdot D \slashed{u} \chi_{v}(x) +  \bar H^{(1)}_{{\tilde v}} (x)   (2 E + i {\tilde  v} \cdot D ) \slashed{u}  H^{(1)}_{{\tilde v}}(x) \right. \\
\nonumber
& +  \left. \bar \chi_{v} (x) ( i \slashed{D}_\perp -m)   H^{(1)}_{{\tilde v}}(x) +  \bar H^{(1)}_{{\tilde v}} (x) (  i \slashed{D}_\perp -m)   \chi_{v}(x) \right] \ ,
\label{initialL}
\end{align}
where we have kept only the terms of the Lagrangian which respect energy-momentum conservation, and
we have defined
$\slashed{D}_{\perp} = P^{\mu \nu}_{\perp} \gamma_\mu D_\nu$
and
\be
\label{transverse-projector}
P^{\mu \nu}_{\perp} = g^{\mu \nu} - \frac 12 \left( v^\mu {\tilde v}^\nu +v^\nu {\tilde v}^\mu\right) \ .
\ee
With our conventions, note that $k^2_\perp = P^{\mu \nu}_{\perp} k_\mu k_\nu = - {\bf k}^2_\perp$.

Using the classical equations of motion one finds  
\be
 H^{(1)}_{{\tilde v}} (x) =  -  \slashed{u} \frac{(i\slashed{D}_\perp -m)}{2 E+ i {\tilde  v} \cdot D } \chi_{v}(x) =  \frac{(i\slashed{D}_\perp +m)}{2 E+ i {\tilde  v} \cdot D }  \frac{ \slashed{\tilde v}}{2} \chi_{v} (x) \ , \label{eq:Hfield}
\ee
where we used that $\slashed{v} \chi_{v} (x)=0 $.

It is not difficult to reach to the modification of the OSEFT Lagrangian. In a general frame is then written down as 
\begin{eqnarray}
\mathcal{L}_{E, v}& = &
\bar \chi_{v}(x) \left(i\, v\cdot D\,
+i \slashed{D}_{\perp}  \frac{1}{2 E + i \tilde{v}\cdot D   }i \slashed{D}_{\perp}  -m^2 
 \frac{1}{2 E + i \tilde{v}\cdot D   }
\right)  \frac{ \slashed{\tilde v} }{2}  \chi_{v}(x) 
\nonumber \\
&-& \bar \chi_{v}(x) \left(  m \Big [ \frac{1}{2 E + i \tilde{v}\cdot D   },   i \slashed{D}_{\perp}   \Big  ]
\right)\frac{ \slashed{\tilde v} }{2}  \chi_{v}(x) 
\, .\label{Leff-SCET-chiral}
\end{eqnarray}
for particles.  We note that the the Lagrangian above results very similar to that
of SCET with a small mass~\cite{hep-ph/0505030}, taking into account the differences between these two effective field theories, that were
already discussed in Ref.~\cite{Carignano:2018gqt}.

The Lagrangian can be now expanded using that $E$ is the hard scale of the problem. We thus write
\be
\mathcal{L}_{E, v} = \sum_{n=0}^\infty \bar \chi_{v}(x) \ {\cal O}_x^{(n)} \ \chi_{v}(x) \ ,
\ee 
where 
\begin{align}
{\cal O}_x^{(0)} &= i\, v\cdot D \, \frac{\slashed{\tilde{v}}}{2} \ , \\
{\cal O}_x^{(1)} &= - \frac {1}{2 E} \,\left( D_{\perp}^{2} + m^2 + \frac e2 \sigma^{\mu \nu}_\perp F_{\mu \nu} \right) \frac{\slashed{\tilde{v}}}{2}\, \ , \\
{\cal O}_x^{(2)} &=   -\frac{1}{4 E^{2}}  (i \slashed{D}_\perp -m)  (i {\tilde v} \cdot D)(i \slashed{D}_\perp +m) \frac{\slashed{\tilde{v}}}{2} 
\label{n=2Lag}
\\
\nonumber
&= \frac{1}{8E^{2}}  \Big(\left[ \slashed{D}_{\perp}\,,\,\left[i\tilde{v}\cdot D\,,\,\slashed{D}_{\perp}\right]\right] +
\left\{ (  \slashed{D}_\perp)^2 + m^2 , i\tilde{v}\cdot D \right\}  \Big)  \frac{\slashed{\tilde{v}}}{2}   - \frac{iem}{4E^2} \tilde {v}^\mu F_{\mu \alpha} \gamma^\alpha_\perp  \frac{\slashed{\tilde{v}}}{2}
\,\ , \nonumber
\end{align}
and so on. We limit our study to operators up to $1/E^2$ in the energy expansion. 
Note that at orders $n=0,1$ chirality is respected. However, at order $n=2$ there is one operator linear in $m$ that breaks chirality.

It is  convenient to introduce local field redefinitions to eliminate the temporal derivative in Eq.~(\ref{n=2Lag}), as in Ref.~\cite{Manuel:2016wqs}. These simplify the computations at higher orders\footnote{Recall that local field redefinitions might not be respectful of the RI if one considers off-shell quantities, but they will not affect the result of on-shell quantities~\cite{Arzt:1993gz}.}.
Thus, after the field redefinition
 %\noindent 
\begin{equation}
\label{LFR}
\chi_{v}\rightarrow\chi_{v}^{\prime}=\left(1 +\frac{   \slashed{D}_{\perp}^{2}+ m^2}{8E^{2}}\right)\chi_{v}\ , 
\end{equation}
the second-order differential operator becomes
\be
{\cal O}_{x, \rm LFR}^{(2)} =  \frac{1}{8E^{2}}  \Big(\left[ \slashed{D}_{\perp}\,,\,\left[i\tilde{v}\cdot D\,,\,\slashed{D}_{\perp}\right]\right] -
\left\{ (  \slashed{D}_\perp)^2+ m^2 ,\,\left(iv\cdot D-i\tilde{v}\cdot D\right)\right\} + 2iem \tilde {v}^\alpha F_{\mu \alpha} \gamma_\perp^\mu \Big)  \frac{\slashed{\tilde{v}}}{2} \ . 
\ee
This is the operator that we will use in our computations of both the dispersion relations and transport equations.

An important observation for our further developments is that we assume that the underlying system is
isotropic. After the split of momenta is done in Eq.~(\ref{eq:Qpart}), and summing over all possible velocities in the OSEFT Lagrangian, we
assume that there is no preferred direction in the system. It would be very interesting to generalize the OSEFT
for anisotropic systems, but this is out of the scope of this project.

%%%%%%%%%%%%%%%%%%%%%%%%%%%%%%%%%%%%%%%%%
\section{Fermion Wigner function}
\label{WigFun}
%%%%%%%%%%%%%%%%%%%%%%%%%%%%%%%%%%%%%%%%%%%

We define the Wigner function as in Ref.~\cite{Carignano:2018gqt}
\be
\label{progaKS}
S_{E,v} (x,y) = 
\left( \begin{array}{cc}
  S^{c}_{E,v} (x,y) & S^<_{E,v} (x,y) \\
 S^>_{E,v}(x,y) & S^{a}_{E,v}(x,y) \\
\end{array} 
\right ) =
\left( \begin{array}{cc}
\langle T  \chi_v(x) \bar \chi_v(y) \rangle & - \langle \bar \chi_v(y)  \chi_v(x) \rangle \\
\langle   \chi_v(x) \bar \chi_v(y) \rangle & \langle {\tilde T}  \chi_v(x) \bar \chi_v(y) \rangle 
\end{array} 
\right ) \ ,
\ee
where $T$ denotes time ordering operator, and ${\tilde T}$ anti-time ordering operator. To ease the notation we removed the prime in our fermion fields, but work with the fermion fields after the local field redefinition of Eq.~(\ref{LFR}) has been performed.

We pay now attention to the different possible structures in Dirac space of the two-point functions, considering the constraints imposed in the OSEFT.
One can  expect that the two-point functions have scalar, vectorial and tensorial components. If we write
\be
\label{Dirac-decomposition}
 S^<_{E,v} = \sum_{\chi= \pm}  P_\chi  \left({\cal F}_S^\chi + \gamma_\mu   J ^{\mu,\chi}   + i \frac{\chi}{4} (G_T^\chi)^{\mu \nu} \sigma_{\mu \nu}\right)\ ,
\ee
where $\sigma_{\mu\nu}=i[\gamma^\mu,\gamma^\nu]/2$, $\chi =\pm$ is an index that indicates the chirality of the particle, and where 
\be
P_\chi = \frac{( 1 + \chi \gamma_5)}{2} \ , 
\ee
 is a chirality projector. We will also use the notation $R$ for $\chi=+$, and $L$ for $\chi=-$. The hermiticity condition of $S^<_{E,v}$ imposes that
${\cal F}_S^R = ({\cal F}_S^L)^\dagger$, $(J^{\mu,\chi})^\dagger= J^{\mu,\chi}$ and $((G_T^L)^{\mu \nu})^\dagger = (G_T^R)^{\mu \nu}$. In general, one can write the tensorial components without the chirality projector, as $\gamma_5 \sigma_{\mu \nu}$ is not an independent term, but it will turn out to be very convenient to write it down like this, as we will show.

We will study now the constraints imposed in the OSEFT to every component.
In Ref.~\cite{Carignano:2018gqt}  we worked out in detail the constraints obeyed by the vectorial components when $m=0$: in the OSEFT we found that $J^{\mu,\chi}= v^\mu G_{E,v}^{\chi}$, and derived the corresponding transport equation obeyed by $G_{E,v}^\chi$.

It is easy to understand that in the scalar components in the OSEFT are zero, thus ${\cal F}_S^\chi =0$.
This arises because particles and antiparticles are totally decoupled in the OSEFT, and 
\be \Tr [S^<_{E,v}]= - \Tr \langle \bar \chi_v(y)  \chi_v(x) \rangle = - \Tr \langle \bar \chi_v(y) P_{\tilde v} P_v  \chi_v(x) \rangle =0 \ . \label{eq:noscalar} \ee
Equally for $\Tr [\gamma_5 S^<_{E,v}] =0$.
While in the full theory the scalar components of the Wigner function are not zero, it is possible to express them in terms of the remaining components in OSEFT (see Sec.~\ref{app:scalar})

Let us find some of the constraints obeyed by the tensorial components. The $\sigma_{\mu\nu}$ matrix can be decomposed with help of the transverse projector in its different components. By simply using 
\be
\gamma_\mu = \gamma_\mu^\perp + \frac 12 \left( v_\mu \slashed{\tilde v} + {\tilde v}_\mu \slashed{ v} \right) \ ,
\ee
one can see that
\be
 \langle \bar \chi_v(y) \sigma_{\mu \nu} \chi_v(x) \rangle = \frac 12  v_\mu {\tilde v}^\rho\langle \bar \chi_v(y)  \sigma_{\rho \nu_\perp}\chi_v(x) \rangle - \frac 12 v_\nu {\tilde v}^\rho \langle \bar \chi_v(y)  \sigma_{\rho \mu_\perp} \chi_v(x) \rangle \ ,
\ee
and other possible projections of the sigma matrix would vanish as $\slashed{v} \chi_v =0$, and  also $$\langle \bar \chi_v(y) \sigma^ {\mu \nu}_\perp \chi_v(x) \rangle =0 \ . $$

It is convenient to work in the LRF of the system, where $u^\mu =(1,0,0,0)$. Let us introduce two unitary vectors in the transverse directions to $v^\mu$: $n_{\perp,1}^\mu$ and $n_{\perp,2}^\mu$, such that $n_{\perp,i} \cdot n_{\perp,j} = -\delta_{ij}$.
For example, if we take $v^\mu = (1,0,0,1)$ and ${\tilde v}^\mu = (1,0,0,-1)$, then $n^\mu_{\perp,1}= (0,1,0,0)$ and $n^\mu_{\perp,2}= (0,0,1,0)$, such that
$({\bf v} ,{\bf n}_{\perp,1}, {\bf n}_{\perp,2})$ form an orthonormal basis of three dimensional vectors. 
 If we choose ${\bf v}$ in a different direction, the remaining vectors can be obtained simply by rotating them accordingly.

Therefore, we can decompose
\be
 (G_T^\chi)^{\mu \nu} = G_T^{\chi,1} (v \wedge n_{\perp,1})^{\mu \nu}   +  G_T^{\chi,2} ( v  \wedge n_{\perp,2})^{\mu \nu}  \ ,
\label{eq:GT} \ee
where we have introduced the wedge product
$$(U \wedge V)^{\mu \nu} \equiv U^\mu V^\nu - U^\nu V^\mu \  .$$

 We define the dual of a tensor $T_{\mu \nu}$  as $T^\star_{\mu \nu} = \frac i2 \epsilon_{\mu \nu \alpha \beta} T^{\alpha \beta}$. Then  $\sigma_{\mu \nu} P_{R/L} T_{\alpha \beta} $ projects the self-dual/anti self-dual parts of 
$T_{\alpha \beta}$, respectively. Thus,
\begin{align}
{\rm Tr} ( \sigma^{\mu \nu} P_{R} S^<_{E,v} )& =  i \left[ (G_T^R )^{\mu \nu} + ((G_T^R)^{\mu \nu})^\star \right]  \ , \\
{\rm Tr} ( \sigma^{\mu \nu} P_{L} S^<_{E,v} )&=  - i \left[ (G_T^L )^{\mu \nu} - ((G_T^L)^{\mu \nu})^\star \right] \ .
\end{align}

If we define the vectors $n_{\perp,\pm} \equiv n_{\perp,1} \pm i n_{\perp,2}$, then $n_{\perp,+} \cdot n_{\perp,-} = -2$, but $n_{\perp,\pm} \cdot n_{\perp,\pm} = 0$ . 
In this case one finds that  
\be \label{eq:selfdual}
(v \wedge n_{\perp,\pm})^{\mu \nu} = \pm (( v \wedge n_{\perp,\pm})^{\mu \nu} )^\star
\ee
are tensorial combinations which are either self-dual or anti self-dual.

The projections of the Wigner function necessary to obtain the tensorial components can then be written as
\begin{align}
{\rm Tr} ( \sigma^{\mu \nu} P_{R} S^<_{E,v} )& = i  (G^{R,1}_T -i G^{R,2}_T) (v \wedge n_{\perp,+})^{\mu \nu}  \equiv 2 i \Phi_{E,v}^\dagger (v \wedge n_{\perp,+})^{\mu \nu} 
 \ , \label{eq:tracesigma1} \\
{\rm Tr} ( \sigma ^{\mu \nu} P_{L} S^<_{E,v} )& =  - i (G^{L,1}_T + i G^{L,2}_T) (v \wedge n_{\perp,-})^{\mu \nu} \equiv -2 i \Phi_{E,v}( v \wedge n_{\perp,-})^{\mu \nu} \ ,  \label{eq:tracesigma2}
\end{align}
in terms of the so called {\it spin coherence function} $\Phi_{E,v}(x,y)$, introduced in Ref.~\cite{arXiv:1605.09383}, see also \cite{Vlasenko:2013fja}.

As in Ref.~\cite{arXiv:1605.09383} (but without a phase factor, which is a matter of conventions) we can write 
\ba
G^{\chi}_{E,v} (x,y) &= & \frac 12  {\rm Tr} \left( \frac{\slashed{\tilde v}}{2} P_{\chi} S_{E,v} (x,y) \right) 
\label{eq:G}
\ , \\
\Phi^{\chi}_{E,v} (x,y)  & = &\chi \frac{i}{16} {\rm Tr} ( \sigma_{\mu \nu} P_\chi S_{E,v} (x,y) ) ({\tilde v} \wedge n_{\perp,-\chi})^{\mu \nu} \ 
. \label{eq:Phi}
\ea
where, to simplify notation, we also use chirality indices $R/L$ (or $+/-$) for the complex function of the tensor component, in such a way that 
\be
\Phi_{E,v}^- \equiv \Phi_{E,v}^L= \Phi_{E,v} \ , \qquad
\Phi_{E,v}^+ \equiv \Phi_{E,v}^R= \Phi_{E,v}^\dag \ .
\ee

As in Ref.~\cite{arXiv:1605.09383}, it turns out convenient to collect the different components in a $2 \times 2$ matrix
\be
\label{G-matrix}
\hat{G}_{E,v} =  
\begin{pmatrix}
G_{E,v}^L & \Phi_{E,v} \\
\Phi^\dag_{E,v} & G_{E,v}^R 
\end{pmatrix}  , 
\ee
where the diagonal entries, corresponding to the vectorial components, describe pure left and right chiral states, while the off-diagonal entries, those associated to the tensorial components, are those corresponding to the quantum coherent left-right or right-left states.

In this paper we will study how the mass corrections affect the transport equation for the vectorial component, and also the equations for the tensorial component, or spin coherence function. Let us recall here that the antiparticles can de described in analogous terms, so that is why we will focus on the particles in this manuscript, noting that the antiparticles can be treated equally, after minor changes.

%%%%%%%%%%%%%%%%%%%%%%%%
\section{ Dispersion relation and validity of the approach}
\label{Disp-sec}
%%%%%%%%%%%%%%%%%%%%%%%%%%

We start computing the dispersion relation of the vector/axial and tensor components of the Wigner function. The fermion dispersion relation can be determined by computing the spectral function of the system, 
\be \rho(x,y) = i [S_{E,v}^< (x,y) - S_{E,v}^> (x,y)]  \ . \label{eq:spectral} \ee

The spectral function is a matrix in Dirac space, and has antihermiticity properties,
\be \gamma^0 \rho(x,y)^\dag \gamma^0 = -\rho(y,x) \ . \label{eq:antiherm} \ee
Given the definition (\ref{eq:spectral}) it is clear that the spectral function admits a similar expansion as in Eq.~(\ref{Dirac-decomposition}),
\be
\rho = i \sum_{\chi= \pm}  P_\chi  \left( \rho_S^\chi + \gamma_\mu  \rho_V^{\mu,\chi}   +i  \frac{\chi}{4} (\rho_T^\chi)^{\mu \nu} \sigma_{\mu \nu}\right)\ ,
\ee
where the different components must satisfy hermiticity conditions so that (\ref{eq:antiherm}) holds.
In OSEFT, the scalar component is zero, while  the vectorial and tensorial components are limited to the same projections that were found for 
$S^<_{E,v}$.

The equation of motion for $\rho(x,y)$ follows from the equations for $S_{E,v}^{\lessgtr} (x,y)$. A constraint equation can be found as
\be
\label{eq:eqrho}
\left[{\cal O}_x  \, \rho (x,y) + \rho (x,y)U(x,y)\gamma_0 ( {\cal O}_y)^\dag \gamma_0 \right] = 0 \ ,
\ee
which can be used to find the fermion dispersion relation.

We perform a (gauge covariant) Wigner transformation
\be
\label{I*-functions}
 I_{+} = \sum_{n=0}^\infty I_+^{(n)}= \sum_{n=0}^\infty \int d^4s e^{i k\cdot s} \left({\cal O}^{(n)}_x  U(x,y) \, \rho (x,y) + \rho(x,y)U(x,y) \gamma_0{\cal O}^{(n) \dag}_y \gamma_0 \right) \ , \ee
where $U$ is the Wilson link between $x$ and $y$ (see Appendix~\ref{app:Wigner}). Upon application of a gradient expansion, we study the resulting constraint equations in an expansion of $1/E$ as explained in Appendix~\ref{app:details}.

At every order in the OSEFT Lagrangian we find operators with  the Dirac structures
\be
{\cal O}^{(n)}_x = \left( \alpha_x^{(n)} + \beta^{(n)}_{x,\mu\nu}
\sigma_\perp^{\mu\nu}+ \delta_{x,\mu}^{(n)} \gamma^\mu_\perp \right) \frac{\slashed{\tilde v}}{2} \ .
\ee

We display in Table~\ref{tab:ope} the different operators found up to second order in the energy expansion, noting the operators that break chirality are only
those described by the $\delta_{x,\mu}^{(n)}$ pieces, and as already stated, these only start appearing at order $1/E^2$ and beyond.

\begin{table}[ht]
\begin{tabular}{|c |ccc|}
\hline
 & $n=0$ & $n=1$ & $n=2$ \\
 \hline 
 \hspace{5mm}  $\alpha_x^{(n)}$ \hspace{5mm}  & \hspace{5mm} $i v\cdot D_x$ \hspace{5mm} & $-\frac{1}{2E} (D_{\perp,x}^2+m^2)$\hspace{5mm}  & \hspace{5mm} $\frac{1}{4E^2} (v^\alpha - \tilde{v}^\alpha) [eF_{\mu \alpha}(x) D_x^\mu -iD_{\alpha,x} (D^2_{\perp,x} +m^2)]$ \hspace{5mm} \\
  $\beta_{x,\mu\nu}^{(n)}$ &0& $-\frac{e}{4E} F_{\mu \nu}(x)$ &  $\frac{i}{4E^2} \left[ eF_{\mu\alpha}(x) \tilde{v}^\alpha D_{\nu,x} - \frac{e}{2} F_{\mu\nu}(x) (v \cdot D_x - \tilde{v} \cdot D_x) \right]$\\
  $\delta_{x,\mu}^{(n)}$ & 0& 0& $i\frac{me}{4 E^2} \tilde{v}^\alpha F_{\mu \alpha} (x)$ \\
  \hline
\end{tabular}
\centering
\caption{Operators appearing in the kinetic equation and dispersion relation at different orders in the power counting.}
\label{tab:ope}
\end{table}
 
It is convenient to take the Dirac trace to distinguish chiralities
\be \textrm{ Tr } I_+ = \sum_\chi I_+^{\chi} = \sum_\chi \sum_{n=0}^\infty I_+^{(n),\chi} \ . \ee
In particular, associated to a given chirality $\chi$, we find 
\begin{align}
  \sum_{n=0}^{\infty}  I_+^{(n),\chi} = \sum_{n=0}^{\infty} \int d^4s e^{i k\cdot s} &\left\lbrace 
 \left(\alpha_x^{(n)} + \alpha_y^{(n)*} \right)  \Tr \left[\frac{\vts}{2} P_\chi
\rho (x,y)\right] \right. \nonumber \\ 
 +& \left. \left(\beta_{x,\mu\nu}^{(n)} +\beta_{y,\mu\nu}^{(n)*} \right) \Tr
\left[ \sigma^{\mu\nu}_\perp \frac{\vts}{2} P_\chi \rho (x,y) \right] \right.
\nonumber \\
+ &\left. \left( \delta_{x,\mu}^{(n)}  - \delta_{y,\mu}^{(n)^*} \right)  \Tr
\left[ \gamma^\mu_\perp \frac{\vts}{2} P_\chi \rho (x,y) \right] \right\rbrace \ .
\label{eq:Iplus}
\end{align}

Similarly to the case for $S^<_{E,v}$ in (\ref{eq:Phi}) we define the scalars functions associated to the vectorial and tensorial components
\begin{align}
 \rho_G^\chi (x,y) &=  \frac 12  \Tr \left[ \frac{\slashed{\tilde v}}{2} P_{\chi} \rho(x,y) \right] \ , \\
\rho_\Phi^\chi(x,y)  & = \chi \frac{i}{16} \Tr \left[ \sigma_{\mu \nu} P_\chi \rho(x,y) \right] ({\tilde v} \wedge n_{\perp,-\chi})^{\mu \nu} \ ,
\end{align}
respectively.

An explicit computation of the Dirac traces and Wigner transformation, together with a gradient expansion gives 
\be I_+^{(0),\chi} + I_+^{(1),\chi} + I_+^{(2),\chi}= 0 \ , \ee
up to order $n=2$, where these functions are given in App.~\ref{app:details}. From that appendix one obtains,
\be
\label{disp-spectral}
2  K_m^\chi \rho_G^\chi (X,k)  + \chi i B_m^\chi \rho_\Phi^\chi (X,k) = 0 \ ,
\ee
where 
\ba
\label{dispersion-n2}
K_m^\chi & \equiv &  2 k \cdot v + \frac{1}{E}  \left(  k^2_{\perp} - m^2 - \frac{e \chi}{4} \epsilon^{\alpha \beta \mu_\perp \nu_\perp}  {\tilde v}_\beta v_{\alpha} F_{\mu \nu}
  \right)   \\
  & - &  \frac{1}{E^2} \left[ \left(  k^2_{\perp}   - m^2- \frac{e \chi}{4} \epsilon^{\alpha \beta \mu_\perp \nu_\perp}  {\tilde v}_\beta v_{\alpha} F_{\mu \nu} \right)  \frac{{\tilde v}\cdot k - v \cdot k}{2}  + \frac{e \chi}{4}     \epsilon^{\alpha \beta \mu_\perp \nu_\perp} 
   {\tilde v}_\beta v_{\alpha} F_{\nu \rho} {\tilde v}^\rho k_{\mu}   \right] \nn 
\ea
and it agrees with the function $K^\chi$ of Ref.~\cite{Carignano:2018gqt}, but here small mass corrections are included. 
We have also defined the function
\be B_m^\chi \equiv \frac{e m}{E^2}  {\tilde v}^\mu F_{\mu \alpha} n^\alpha_{\perp, \chi} \ , \ee
to ease the notation.

On the other hand, if we multiply Eq.~(\ref{I*-functions}) by $\slashed{n}_{\perp,\chi}$ and take the trace, we end up with an independent equation relating the vectorial and tensorial components of the spectral function. Defining
\be I_{\Phi^\chi;+} \equiv \Tr(\slashed{n}_{\perp,\chi}I_+) = \sum_{n=0}^\infty I_{\Phi^\chi;+}^{(n)} \ , \label{eq:IPhiplus} \ee
we have
\ba
0 &=& I_{\Phi^\chi;+} (x,y) = \sum_{n=0}^\infty \int d^4s \, e^{i k\cdot s} \left\lbrace
\left(\alpha_x^{(n)} + \alpha_y^{(n)*} \right)  \Tr \left[\slashed{n}_{\perp,\chi}\frac{\slashed{\tilde v} }{2}
\rho (x,y)\right]  \right. 
 \\ &+&
\left.
\left(\beta_{x,\mu\nu}^{(n)} + \beta_{y,\mu\nu}^{(n)*} \right) \Tr
\left[\slashed{n}_{\perp,\chi}  \sigma^{\mu\nu}_\perp \frac{\slashed{\tilde v}}{2} \rho(x,y) \right]
% \right. 
\nonumber
 + 
( \delta_{x,\mu}^{(n)}  - \delta_{y,\mu}^{(n)^*})  \Tr
\left[\slashed{n}_{\perp,\chi} \gamma^\mu_\perp \frac{\vts}{2} \rho(x,y) \right] \right\rbrace 
 \ea

As before, we will keep terms up to $n=2$,
\be 0 = I^{(0)}_{\Phi^\chi;+}+I^{(1)}_{\Phi^\chi;+}+I^{(2)}_{\Phi^\chi;+}  \ . \ee 
An explicit computation of the above traces, and performing the gradient expansion together with the Wigner transform (see App.~\ref{app:details}) gives
\begin{align}
\label{relation-Phi}
 0 &=  4 \chi \left\{ 2 k \cdot v + \frac{1}{E} \left[ k^2_\perp -m^2- i \chi \frac e4 ( n_{\perp,+} \wedge n_{\perp,-} )^{\mu \nu} F_{\mu\nu} (X) \right] \right.
\\
  & -  \left. \frac{1}{E^2} \left[ \left(  k^2_{\perp}   - m^2-  i  \chi \frac e4  (n_{\perp,+} \wedge n_{\perp,-} )^{\mu \nu} F_{\mu\nu} (X) \right)  \frac{{\tilde v}\cdot k - v \cdot k}{2} + i\chi \frac e4
 ( n_{\perp,+} \wedge n_{\perp,-} )^{\mu\nu} F_{\nu \rho} (X){\tilde v}^\rho k_{\mu}   \right] \right\} \rho_\Phi^\chi (X,k)
 \nonumber   \\
 &- i\frac{2me}{E^2} \tilde{v}^\mu F_{\mu\alpha} (X)n_{\perp,-\chi}^\alpha \rho_G^{-\chi} (X,k)  \ . \nn
\end{align}

At this point we make the following observation: the dispersion relation and the transport equation for $\Phi_v^\chi$ can be written down in a form
quite parallel to that for $G_{E,v}^\chi$. First note that
\be
  \frac{i}{2} e F_{\mu \nu} (X) (n_{\perp,+} \wedge n_{\perp,-})^{\mu \nu}=   e  F_{\mu \nu} (X) ( n_{\perp,1} \wedge n_{\perp,2})^{\mu \nu} \ .
\ee
In the LRF, where $u^\mu = (1,0,0,0)$, and taking into account that the unitary vectors $({\bf v} ,{\bf n}_{\perp,1}, {\bf n}_{\perp,2})$ are unitary and orthonormal, it is possible to express 
\be \label{eq:useful}
( n_{\perp,1} \wedge n_{\perp,2})^{\mu \nu} = \epsilon^{\mu \nu \alpha \beta} v_\alpha u_\beta  \ .
\ee
The same relation holds in any frame, noting that $(u,v,n_{\perp,1}, n_{\perp,2})$ form a basis in Minkowski space, and multiplying the two sides of the equation by any combination of two of these four vectors gives the same result.

We then find that the constraint equation for the tensorial component (\ref{relation-Phi}) can be expressed as
\be 2 \chi K_m^\chi \rho_\Phi^\chi (X,k) 
 - iB_m^{-\chi} \rho_G^{-\chi} (X,k) = 0 \ . \label{eq:relation-Phisimp}  \ee

From Eqs.~(\ref{eq:relation-Phisimp}) and (\ref{disp-spectral}) one can see that the four components $\rho_G^R,\rho_G^L,\rho_\Phi^R,\rho_\Phi^L$ are mixed. Let us write explicitly the four independent equations making use of the relation (\ref{eq:useful}) to be able to express Eq.~(\ref{relation-Phi}) in terms of the $K_m^\chi$ given in (\ref{dispersion-n2}). 

It is easy to partially decouple the system of equations 
\be 
\left\{ 
\begin{array}{ccc}
4  K_m^\chi K_m^\chi \rho_G^\chi (X,k)  - B_m^\chi B_m^{-\chi}  \ \rho_G^{-\chi} (X,k)  & =& 0 \ , \\
4  K_m^\chi K_m^{-\chi} \rho_\Phi^\chi (X,k)  + B_m^{-\chi} B_m^{-\chi}  \ \rho_\Phi^{-\chi} (X,k)   &= &0  \ . 
\end{array}
\right.
\ee
Thus, the vectorial and tensorial components are decoupled, while the two chiralities of every component are coupled, but this only happens at order  ${\cal O}(1/E^4)$.

If we work at lower orders, as we do in this manuscript, we can neglect that coupling, and then write
\be
\left\{
\begin{array}{ccc}
 K_m^\chi \rho_G^\chi (X,k)    & = &0 + {\cal O} (1/E^3)  \ ,  \\
 K_m^\chi \rho_\Phi^\chi (X,k) & =&0 + {\cal O} (1/E^3)  \ . 
\end{array}
\right.
\ee
so that
\begin{align}
\rho_G^\chi (X,k)  &  = 2\pi  \delta(K_m^\chi) + {\cal O} (1/E^3)  \ ,  \\
\rho_\Phi^\chi (X,k) & =  2\pi  \delta(K_m^\chi) + {\cal O}(1/E^3)  \ ,
\end{align}
where the factor $2\pi$ is simply added for convenience.

It is not difficult to see that, if we return to the full variables the argument of the Dirac delta function describes the dispersion law of 
 \be
\label{eq:onshell}
q^2  - m^2- e S_\chi^{\mu \nu} F_{\mu\nu}(X) = 0  \, ,
\ee
for positive values of  $E_q= q \cdot u$, the energy variable in the local rest frame, and
where $S_\chi^{\mu \nu} $ is the spin tensor defined as 
\be
S_\chi^{\mu \nu} =  \chi\frac{ \epsilon^{\alpha \beta \mu \nu} u_\beta q_\alpha}{2 (q \cdot u)}  \ ,
\ee
when expanded in powers of $1/E$ up to second order (see Ref.~\cite{Carignano:2018gqt}), but correctly including here small mass corrections.

\section{Derivation of the collisionless transport equations with small mass corrections}
\label{Derivation}

In this section we generalize our work of Ref.~\cite{Carignano:2018gqt} to see how the chiral kinetic equation for the vectorial components is modified by the presence of a small mass, and we further derive the transport equation associated to the tensorial components.

We start by considering the equations obeyed by the two-point Green's functions, as follows from the OSEFT Lagrangian. To derive the collisionless transport equation it is enough to consider the  tree level equations. We then consider the combination
\be
\label{I-functions}
 I_{-} =  \int d^4s e^{i k\cdot s} \left({\cal O}_x  U(x,y) \, S^<_{E,v} (x,y) - S^<_{E,v} (x,y)U(x,y) \gamma_0 {{\cal O}^ \dag}_y \gamma_0  \right) \ , \ee
where $U$ is the Wilson link between $x$ and $y$, needed to render the Wigner function gauge invariant (see App.~\ref{app:Wigner}).

In order to recover the transport equation associated to the vectorial components of the two-point function one simply has to take the trace to the above equations
\be \textrm{ Tr } I_-^{(n)} = \sum_\chi I_-^{(n),\chi} \ . \ee
In particular, associated to a given chirality $\chi$, we find 
\begin{align}
   I_-^{(n),\chi} = \int d^4s e^{i k\cdot s} &\left\lbrace 
 \left(\alpha_x^{(n)} -\alpha_y^{(n)*} \right)  \Tr \left[\frac{\vts}{2} P_\chi
S_{E,v} (x,y)\right] \right. \nonumber \\ 
 +& \left. \left(\beta_{x,\mu\nu}^{(n)} - \beta_{y,\mu\nu}^{(n)*} \right) \Tr
\left[ \sigma^{\mu\nu}_\perp \frac{\vts}{2} P_\chi S_{E,v}(x,y) \right] \right.
\nonumber \\
+ &\left. \left( \delta_{x,\mu}^{(n)}  + \delta_{y,\mu}^{(n)^*} \right)  \Tr
\left[ \gamma^\mu_\perp \frac{\vts}{2} P_\chi S_{E,v}(x,y) \right] \right\rbrace \ .  \label{eq:Iminus}
\end{align}

While the first two terms select the vectorial components of the two-point function,  
the trace in the last term of the above equation turns out to be proportional to $\Phi$, the tensorial component. 
However, because the $\delta_\mu$ functions are pure imaginary (see Table~\ref{tab:ope}), such a term cancels, and thus the tensorial components  do not affect the transport equation for
vectorial components.
The mass terms in the OSEFT Lagrangian only provide corrections of the dispersion law (see Appendix~\ref{app:details} for details), and thus the corresponding on-shell velocity, at the desired order of accuracy.

In the $I_{-}^{\chi}$ functions we then reproduce the velocity vector of an energetic particle with a small mass. 
If we now  include the gauge fields, we then get the mass correction to the on-shell velocity we found in Ref.~\cite{Carignano:2018gqt}
\be
\label{onshellvector}
v_\mu^\chi \Big |_{\rm o.s.} \equiv \frac{q^\mu}{E_q}  \Big |_{\rm o.s.} = v^\mu +  \frac{k^\mu_\perp}{E} - (k \cdot {\tilde v}) \frac{k^\mu_\perp}{2E^2} +\frac{v^\mu -{\tilde v}^\mu}{4 E^2} \left[ k^2_\perp - m^2 -
\frac{e \chi}{4} \epsilon^{\alpha \beta \mu \nu} {\tilde v}_\beta v_{\alpha} F^\perp_{\mu \nu} (X) \right]
 + {\cal O}\left(\frac{1}{E^3}\right) \ 
\ee
where $E_q  = q \cdot u$. This is the combination that appears in the $I_{\chi,-}$ functions (see App.~\ref{app:details}).

 Similar to the $1/E$ expansion of the on-shell velocity, it is convenient to define the rest of the orthogonal basis as long as one proceed with the OSEFT expansion. The expansion of $\tilde{v}$ is trivial as they are related by the frame vector $u^\mu$ which does not depend in the order of the expansion. In the on-shell case we will always replace $\tilde{v}^\mu$ by $2u^\mu-v_\chi^\mu$. Then the perpendicular vector should also be modified, at order $1/E$ we define 
\be n^\mu_{\perp,\chi,q} \equiv n^\mu_{\perp,\chi} - \frac{k_\perp \cdot n_\perp^\chi}{2E}\tilde{v}^\mu + \frac{k_\perp \cdot n_{\perp}^\chi}{2E} v^\mu \ , \label{eq:nperpq} \ee
so that it is orthogonal to both $v_\chi^\mu$ and $\tilde{v}^\mu$ at ${\cal O}(1/E)$.

From
\be I_-^{(0),\chi} + I_-^{(1),\chi} +I_-^{(2),\chi} =0\ee
after Wigner transform we obtain
\be 2i\left[ v^\mu_\chi - \frac{e\chi}{8 E^2} \epsilon^{\alpha \beta \mu_\perp \nu_\perp} \tilde{v}_\beta v_\alpha F_{\nu \rho} (X) \tilde{v}^\rho \right] [ \pa_{X,\mu}  -e  F_{\mu \nu} (X) \pa_k^\nu ] G_{E,v}^\chi(X,k)   = 0 \ .\ee

In order to obtain the transport equations and dispersion law for $\Phi_{E,v}^\chi$ one has to multiply Eq.~(\ref{I-functions}) by $\slashed{n}_{\perp,-\chi}$ ,  and then take the trace
\ba
I_{\Phi^\chi;-}^{(n)} &\equiv& \Tr(\slashed{n}_{\perp,-\chi}I_-^{(n)}) = \int d^4s \, e^{i k\cdot s} \left\lbrace
\left(\alpha_x^{(n)} - \alpha_y^{(n)*} \right)  \Tr \left[\slashed{n}_{\perp,-\chi}\frac{\slashed{\tilde v} }{2}
S^<_{E,v} (x,y)\right]  \right. \label{eq:IPhiminus}
 \\ &+&
\left.
\left(\beta_{x,\mu\nu}^{(n)} - \beta_{y,\mu\nu}^{(n)*} \right) \Tr
\left[\slashed{n}_{\perp,-\chi}  \sigma^{\mu\nu}_\perp \frac{\slashed{\tilde v}}{2} S^<_{E,v}(x,y) \right]
% \right. 
\nonumber
 + 
( \delta_{x,\mu}^{(n)}  + \delta_{y,\mu}^{(n)^*})  \Tr
\left[\slashed{n}_{\perp,-\chi} \gamma^\mu_\perp \frac{\vts}{2}  S^<_{E,v}(x,y) \right] \right\rbrace \ .
 \ea
%\right\rbrace \,,

We perform the gradient expansion of the different operators, and find up to second order in the energy expansion
 \begin{align}
   &4 i \chi \left\{  v^\mu+ \frac{k^\mu_\perp}{E} + 
 \frac{1}{E^2} \left( - k^\mu_{\perp} \frac{{\tilde v}\cdot k - v \cdot k}{2}   + \frac{ 1}{4} \left[  k^2_{\perp} -m^2 - i \chi \frac{ e}{4} F_{\alpha \beta} (X) (n_{\perp,+} \wedge n_{\perp,-} )^{\alpha\beta}
 \right] (v^\mu - {\tilde v}^\mu ) 
 \right. \right. \nn \\
 &  \left. \left.  - i \chi \frac e8  ( n_{\perp,+} \wedge n_{\perp,-} )^{\mu\nu}  F_{\nu \rho} (X) {\tilde v}^\rho  
  \right) \right\} [ \pa_{X,\mu}  -e  F_{\mu \nu} (X) \pa_k^\nu ] \ \Phi_{E,v}^\chi (X,k) 
= 0 \ .  \label{transeqk-Phi} 
\end{align}

As in the computation of the spectral function we note that if we use  Eq.~(\ref{eq:useful}), the transport equation for $\Phi_{E,v}^\chi$ can be written in terms  of the on-shell vector velocity (\ref{onshellvector}), and that the transport equations for the tensorial components turn out the same as that for the vectorial components.

If we write
\ba
G^\chi_{E,v}(X,k) &=& \rho^\chi_G(X,k) f^\chi_{E,v} (X,k) =2 \pi \delta(K^\chi_m) f^\chi_{E,v} (X,k) \ , \\
\Phi^\chi_{E,v} (X,k) &=& \rho^\chi_\Phi (X,k) \phi_{E,v}^\chi(X,k) = 2 \pi \delta(K^\chi_m) \phi^\chi_{E,v} (X,k) \ ,
\ea
where $f^\chi_{E,v}$ and $\phi^\chi_{E,v}$ are the distribution functions associated to the vectorial and tensorial components, respectively, but expressed in terms of the residual momenta. One can express the same functions in terms of the full momenta $q$ and write the transport equations as

\begin{align}
 \left[ v^\mu_\chi - \frac{e}{2 E^2_q}S_\chi^{\mu \nu}F_{\nu \rho}  (X) \left(2 u^\rho - v^\rho_\chi \right) \right] \Delta_\mu G^\chi(X,q)  &  = 0 \ , \label{CCTEq-1} \\
 \left[   v^\mu_\chi - \frac{e}{2 E^2_q}S_\chi^{\mu \nu}F_{\nu \rho} (X) \left(2 u^\rho - v^\rho_\chi \right) \right] \Delta_\mu \Phi^\chi(X,q) & = 0  \ , \label{CCTEq-2}
\end{align}
where we have defined the operator
\be
\Delta^\mu \equiv \pa^\mu_X  -e  F^{\mu \nu} (X) \pa_{q, \nu} \ , \label{eq:Delta}
\ee
and the Green's functions are now expressed in terms of the full momenta
\be  G^\chi_{E,v}(X,k) = G^\chi(X,q) \ , \qquad
\Phi^\chi_{E,v}(X,k) = \Phi^\chi(X,q) \ . \ee

If we integrate the transport equation over $q^0$, taking into the account the on-shell condition, the equations take a rather simple form in
the local rest frame $u^\mu =(1,0,0,0)$. Using $F^{i0}= E^i$, $F^{ij} = - \epsilon^{ijk} B^k$, and also that in this frame
\be
S_\chi^{\mu \nu} \rightarrow S^{ij}_\chi = \chi \frac{ \epsilon^{ijk}q^k}{2 q^0} \ , \qquad  S_\chi^{\mu \nu} F_{\mu\nu} = - \chi \, {\bf B} \cdot \frac{\bf q}{q^0}
\ee 

At the order we have worked, and in this frame, one can easily find how the mass term modifies the on-shell vector $v^i_q$ at order $1/q^2$, so that one finds  
for the on-shell distribution function
\ba
\label{staticCKT-1}
\nonumber
 \left ( \Delta_0 + {\bf  \hat q}^i \left (1 - \frac{m^2}{2q^2}+ e \chi \frac{ \bf B \cdot \bf \hat q}{2 q^2} \right) \Delta_i  + e \chi \frac{\epsilon^{ijk} E^j \hat{q}^k - B^i_{\perp,\bf q}}{4 q^2}  \Delta_i \right) f^\chi(X, {\bf q}) = 0 \ , \\
\label{staticCKT-2}
\nonumber
 \left ( \Delta_0 + {\bf  \hat q}^i \left (1 - \frac{m^2}{2q^2}+ e \chi \frac{ \bf B \cdot \bf \hat q}{2 q^2} \right) \Delta_i  + e \chi \frac{\epsilon^{ijk} E^j \hat{q}^k - B^i_{\perp,\bf q}}{4 q^2}  \Delta_i \right) \phi^\chi(X, {\bf q}) = 0 \ ,
\ea
where we have defined $B^i_{\perp,\bf q} \equiv B^i -{\bf \hat q}^i ({\bf B} \cdot {\bf \hat q})$.

Thus, after including mass corrections, we see that the transport equations for the vectorial components are only modified by correcting properly the fermion dispersion law and the associated velocity. And the transport equations for the tensorial components are exactly the same than their vectorial counterparts.

Let us recall that  in the massless case the equations obeyed by the vectorial components as derived from the OSEFT differs from others, derived in a $\hbar$-expansion \cite{Hidaka:2016yjf}.
The reason is that the two equations are valid for different degrees of freedom \cite{Lin:2019ytz}. OSEFT is the quantum field theory counterpart of a Foldy-Wouthuysen diagonalization  \cite{Carignano:2019zsh} and the resulting OSEFT field is a combination of the original Dirac particle and antiparticle fields, but is free of {\it Zitterbewegung} oscillations that mixes up their dynamical
evolution. Thus, we may expect that if similar transport equations are rederived in an $\hbar$ expansion when mass corrections are included, the equations might differ, even if in the pure Dirac treatment a way of eliminating the particle-antiparticle oscillations has to be prescribed.

\section{Reparametrization invariance of OSEFT with a small mass }
\label{RiOSEFT}

In this section we discuss how the presence of a small mass modifies the RI of the OSEFT studied in Ref.~\cite{Carignano:2018gqt}, which is only strictly valid in the massless case.
 These modifications are exactly the same as those
of RI of SCET in the presence of a small mass, that were originally discussed in Ref.~\cite{hep-ph/0505030}, as one can naturally expect. We then derive the
transformation rules of the four Wigner functions in the system  under one of these transformations, and ultimately, of their corresponding distribution functions.

RI is the symmetry associated with the ambiguity of the decomposition of the full momentum $q^\mu$ in Eq.~(\ref{eq:Qpart}). A small shift in
the velocity $v^\mu$ could be reabsorbed in the definition of the residual momentum $k^\mu$, while the physics should remain unchanged. On the other hand, the explicit choices of the vectors $v^\mu$ and $u^\mu$ seem to imply an apparent breaking of the Lorentz symmetry. Checking the RI of the theory ultimately confirms that Lorentz symmetry is respected in the effective field theory.

Let us review how this effectively works~\cite{Carignano:2018gqt}, and what are the modifications needed in the presence of a small mass. The Dirac field defined in Eq.~(\ref{eq:Fields}) should be independent of the choice of the parameters used to define the effective field theory, thus 
\be
\label{eq-RI}
\psi_{ v, \tilde v} (x)= \psi'_{ v', \tilde v'}(x) \ .
\ee

As in SCET~\cite{Manohar:2002fd}, we showed in Ref.~\cite{Carignano:2018gqt} that the effective field theory action  remains invariant under  infinitesimal changes of the vectors $v^\mu$ and ${\tilde v}^\mu$ that preserve their basic properties 
expressed in Eq.~(\ref{light-condition}). There are three types of RI transformations \cite{Manohar:2002fd}, 
which can be understood in terms of combinations of infinitesimal Lorentz boosts and rotations, which are 
\begin{eqnarray}
({\rm I}) \, \Bigg \{ \begin{array}{ccl}
v^\mu &\xrightarrow{I} &v^\mu + \lambda_\perp^\mu  \\
{\tilde v}^\mu & \xrightarrow{I} &{\tilde v}^\mu  
%u^\mu &\rightarrow &u^\mu + \frac{\lambda_\perp^\mu}{2}
\end{array}   \qquad  ({\rm II}) \,  \Bigg \{ \begin{array}{ccl}
v^\mu &\xrightarrow{II} &v^\mu   \\
{\tilde v}^\mu & \xrightarrow{II} &{\tilde v}^\mu + \epsilon_\perp^\mu 
%u^\mu & \rightarrow &u^\mu + \frac{\epsilon_\perp^\mu}{2} 
\end{array} 
  \qquad 
({\rm III}) \, \Bigg \{ \begin{array}{ccl}
v^\mu &\xrightarrow{III} & (1+ \alpha) v^\mu  \\
{\tilde v}^\mu & \xrightarrow{III} & (1-\alpha){\tilde v}^\mu 
%u^\mu  & \rightarrow & u^\mu (1+\alpha) - \alpha {\tilde v}^\mu
  \end{array}  
\end{eqnarray}
where $\{ \lambda_\perp^\mu, \epsilon_\perp^\mu, \alpha\}$ are five infinitesimal parameters, which satisfy $v \cdot \lambda_\perp = v \cdot \epsilon_\perp = {\tilde v} \cdot \lambda_\perp = {\tilde v} \cdot \epsilon_\perp =0$.
The transformation rule of the vector $u^\mu$ can be deduced from 
Eq.~(\ref{eq:uvec}). The transformation rule for the vector $n_{\perp,\chi}^\mu$ can be obtained by requiring that the orthogonality conditions $n_{\perp,\chi} \cdot v = n_{\perp,\chi} \cdot \tilde{v}=0$ hold after the transformation. For example, under type II one finds 
\be n^\mu_{\perp,\chi} \xrightarrow{II}  n^\mu_{\perp,\chi} -\frac12
(\epsilon_\perp \cdot n_{\perp, \chi} ) v^\mu  \ . \label{eq:typeIInperp}\ee
The transformation of other quantities in the massless case are shown in~\cite{Carignano:2018gqt}.

Let us note that in the presence of a small mass the OSEFT Lagrangian Eq.~(\ref{Leff-SCET-chiral}) is not invariant under the type II transformation for the massless case. One can recover invariance if the transformation of the OSEFT field is modified to~\cite{hep-ph/0505030}
\be
\chi_{v} (x) \xrightarrow{II} \left[ 1 +  \frac 12 \slashed{\epsilon}_\perp  \frac{1}{2E + i {\tilde v} \cdot D} \left( i \slashed{D}_\perp-m\right) \right] \chi_{v} (x) \ . 
\ee
This transformation rule is needed to  compensate  changes in the antiparticle part of the full wave function, see Eq.~(\ref{eq:Hfield}), which is however integrated out in the final theory.
Quite interestingly, the above rule implies, when projected over chiralities $\chi_v^\chi (x) \equiv P_\chi \chi_v(x)$, that under a type II transformation in the presence of a small mass term, a change of chirality is induced
\be
\label{TypeII-chiral}
\chi^\chi_{v} (x) \xrightarrow{II} \left(1 +  \frac 12 \slashed{\epsilon}_\perp  \frac{1}{2E + i {\tilde v} \cdot D} \ i \slashed{D}_\perp \right) \chi^\chi_{v} - 
  \frac 12 \slashed{\epsilon}_\perp  \frac{m}{2E + i {\tilde v} \cdot D}  \chi^{-\chi}_{v} \ .
\ee

One can check that after this mass modification the  OSEFT Lagrangian (\ref{Leff-SCET-chiral}) is invariant under these three  RI transformations,
\be
\label{RI-inv-L}
\delta_{({\rm I})} \mathcal{L}_{E, v} = \delta_{({\rm II})} \mathcal{L}_{E, v}  = \delta_{({\rm III})} \mathcal{L}_{E, v} =  0 \ .
\ee

As discussed in Ref.~\cite{Carignano:2018gqt}, once we know how the fields of the OSEFT behave under the three type of RI transformations, we can deduce how the different two-point functions are modified under the same transformations. After performing the (gauge-covariant) Wigner transform, and the gradient expansion, we can derive how the distribution functions behave under these changes. We focus here only on the type II transformations, which are those that imply changes in the distribution functions~\cite{Carignano:2018gqt}.

Using the definitions of the two-point functions, Eqs.~(\ref{eq:G}) and (\ref{eq:Phi}), and going up to order $1/E$, after Wigner transform and the gradient expansion, we obtain the following RI type II transformations of $G^\chi_{E,v} (X,k)$ and $\Phi_{E,v}^\chi (X,k)$  
\begin{align} 
G^\chi_{E,v} & \xrightarrow{II}  G^\chi_{E,v} + \frac{1}{2E} \epsilon_\perp \cdot k G^\chi_{E,v} - \frac{\chi}{8E} \epsilon^{\mu_\perp \nu_\perp \alpha \beta } v_\alpha \tilde{v}_\beta \epsilon^\perp_\nu \Delta_{k,\mu} G_{E,v}^\chi + \frac{\chi m}{4E} \epsilon_\perp \cdot [n_{\perp,+} \Phi_{E,v}^\dag + n_{\perp,-} \Phi_{E,v}] \ , \label{eq:GvtypeII} \\
\Phi_{E,v}^\chi & \xrightarrow{II}  \Phi_{E,v}^\chi + \frac{1}{2E} \epsilon_\perp \cdot k \ \Phi_{E,v}^\chi - \frac{\chi}{4E} \epsilon^{\mu_\perp \nu_\perp \alpha \beta } v_\alpha \tilde{v}_\beta
\epsilon^\perp_\mu (i k_\nu) 
\Phi_{E,v}^\chi - \frac{m}{4E} \epsilon_\perp \cdot   n_{\perp,-\chi} \ (G^R_{E,v} - G^{L}_{E,v} ) \ . \label{eq:PhitypeII}
\end{align}

The transformations above lead us to the following transformation of the distribution functions, which we already write here in terms of the full variables
%\ba
\begin{align}
\label{side-jumpf}
f^\chi(X, q)  &\xrightarrow{II} f ^\chi (X,q) - \frac{1}{E_q}  S^{\mu \nu}_\chi \epsilon_\nu^\perp \Delta_\mu 
f^\chi (X,q) + \frac{\chi m}{4E_q} \epsilon_\perp \cdot [n_{\perp,+,q} \phi^\dag(X,q) + n_{\perp,-,q} \phi(X,q)] \ , \\
\phi^\chi(X, q) & \xrightarrow{II}  \phi^\chi (X,q)- \frac{m}{4E_q} \epsilon_\perp \cdot   n_{\perp,-\chi,q} \ (f^R(X,q) - f^L(X,q) ) \  .
\end{align}
%\ea

We note that while the transformation of the distribution functions associated with the vectorial components describes the so-called side jumps, first discussed in Ref.~\cite{Chen:2015gta}, the distribution function associated to the tensorial components do not exhibit such an effect. On the other hand, in the presence of a small mass the distribution function associated to the vectorial, of pure $RR$ or $LL$ states, transforms into that of tensorial, or coherent states $RL$ and $LR$. This is in concordance to the fact that chirality is not an invariant in the OSEFT, see Eq.~(\ref{TypeII-chiral}).

These transformations are needed to preserve the Lorentz symmetry. For example, we will show in App.~\ref{app:typeIISigma} that the
scalar density behaves as a Lorentz scalar, by checking that it is invariant under the type II RI transformation.

%%%%%%%%%%%%%%%%%%%%%%%%%%%%%
\section{Macroscopic variables of the system}
\label{macro}
%%%%%%%%%%%%%%%%%%%%%%%%%%%%%

In this section we compute the scalar and pseudoscalar fermion densities, as well as the vector and axial 4-currents. We focus on the particle contribution, which  will be written in terms of $G^\chi(X,q)$ and $\Phi^\chi(X,q)$, functions of the full momentum $q$. The antiparticle contribution can be expressed in a similar
way, so we will not explicitly write it down. Intermediate algebraic steps are provided in App.~\ref{app:scalarden} for the case of the scalar density.

%%%%%%%%%%%%%%%%%%%%%
\subsection{QED scalar and pseudoscalar densities in terms of the OSEFT components~\label{app:scalar}}
%%%%%%%%%%%%%%%%%%%%%%

Any two-point function in QED can be decomposed as in Eq.~(\ref{Dirac-decomposition}). We start considering the expressions for the scalar and pseudoscalar densities written in terms of the OSEFT fields,
\ba
\sigma (x) &=& \lim_{y \rightarrow x} \Tr \ \langle \bar \psi(y) \psi(x) \rangle  = \lim_{y \rightarrow x} \Tr \left\langle [ \bar {\chi}_v (y)+ \bar{H}^{(1)}_{\tilde v} (y) ][   {\chi}_v (x)+ H^{(1)}_{\tilde v} (x) ]\right\rangle  \ , \\
j_5 (x) &=& \lim_{y \rightarrow x} \Tr \ \langle \bar \psi(y)\gamma_5 \psi(x) \rangle  = \lim_{y \rightarrow x} \Tr \left\langle [ \bar {\chi}_v (y)+ \bar{H}^{(1)}_{\tilde v} (y) ]\gamma_5[   {\chi}_v (x)+ H^{(1)}_{\tilde v} (x) ]\right\rangle \ ,
\ea
where the point-splitting regularization will help to implement the gradient expansion.

We use now the expressions for $H^{(1)}_{\tilde v}(x)$ in Eq.~(\ref{eq:Hfield}) to expand these densities in a $1/E$ expansion. We will also apply the Wigner transform of all fields and consider the gradient expansion of the different operators. We refer the reader to App.~\ref{app:scalarden} for more details. Up to ${\cal O}(1/E^3)$ we obtain
% \ba
% \rho(X) & = &  \sum_{\chi=\pm} \int \frac{d^4q}{(2 \pi)^4}  \left( \frac {m}{E} G^\chi_{E,v}(X,q)- i \chi \frac{1}{E} (n^\chi_\perp \cdot \Delta) \Phi_{E,v}^\chi (X,q) \right)+  {\cal O} \left( \frac {1}{E^2} \right) \ .,\\
% j_5(X) & = &  \sum_{\chi=\pm} \int \frac{d^4q}{(2 \pi)^4} \left( \frac {m \chi}{E} G^\chi_{E,v}(X,q)-i \frac{1}{E} (n^\chi_\perp \cdot \Delta) \Phi_{E,v}^\chi (X,q) \right ) + {\cal O} \left(\frac {1}{E^2} \right) \ . 
% \ea
\begin{align}
\sigma(X) & =  \sum_{\chi=\pm} \sum_{E,v} \int \frac{d^4k}{(2 \pi)^4}  \left[ \frac {2m}{E} G^\chi_{E,v}(X,k)-  \chi \frac{i}{E} (n^\chi_\perp \cdot \Delta_k) \Phi_{E,v}^\chi (X,k) -\frac{m}{E^2} \tilde{v} \cdot k G^\chi_{E,v}(X,k) \right. \nn \\
&+ \left. \chi \frac{i}{2E^2}   \left( \tilde{v} \cdot k \ n_{\perp}^\chi \cdot \Delta_k + k\cdot n_{\perp}^\chi \ \tilde{v} \cdot \Delta_k - e \tilde{v}^\alpha F_{\alpha\mu} (X) n_{\perp,\chi}^{\mu} \right) \Phi_{E,v}^\chi (X,k) \right] +  {\cal O} \left( \frac {1}{E^3} \right) \ , \label{eq:sigmaresidual} \\
j_5(X) & =   \sum_{\chi=\pm} \sum_{E,v} \int \frac{d^4k}{(2 \pi)^4} \left[ - \frac{i}{E} (n^\chi_\perp \cdot \Delta_k) \Phi_{E,v}^\chi (X,k) + \chi \frac{im}{2E^2} (\tilde{v} \cdot \Delta_k) G^\chi_{E,v} (X,k)\right. \nn \\
& \left.  +\frac{i}{2E^2} \left(\tilde{v} \cdot k \ n^\chi_\perp \cdot \Delta_k + k\cdot n^\chi_\perp \ \tilde{v} \cdot \Delta_k - e\tilde{v}^\alpha F_{\alpha \mu} (X) n^\mu_{\perp,\chi} \right) \Phi_{E,v}^\chi (X,k)\right]+ {\cal O} \left(\frac {1}{E^3} \right) \ , \label{eq:j5residual}
\end{align}
where the operator $\Delta_k^\mu \equiv \pa_X^\mu - e F^{\mu\nu}(X) \pa_{k,\nu}$ is written in terms of the derivative with respect of the residual momentum $k$. We now perform several integrations by parts in the residual momentum. Assuming regular properties at $k=0$ and $k=\infty$ of the distribution functions, the surface terms vanish, and we obtain
\begin{align}
\sigma(X) & =  \sum_{\chi=\pm} \sum_{E,v} \int \frac{d^4k}{(2 \pi)^4}  \left[ \frac {2m}{E} G^\chi_{E,v}(X,k)-  \chi \frac{i}{E} (n^\chi_\perp \cdot \pa) \Phi_{E,v}^\chi (X,k) -\frac{m}{E^2} \tilde{v} \cdot k G^\chi_{E,v}(X,k) \right. \nn \\
&+ \left. \chi \frac{i}{2E^2}   \left( \tilde{v} \cdot k \ n_{\perp}^\chi \cdot \pa + k\cdot n_{\perp}^\chi \ \tilde{v} \cdot \pa - e \tilde{v}^\alpha F_{\alpha\mu} (X) n_{\perp,\chi}^{\mu} \right) \Phi_{E,v}^\chi (X,k) \right] +  {\cal O} \left( \frac {1}{E^3} \right) \ , \label{eq:sigmaresidual2} \\
j_5(X) & =   \sum_{\chi=\pm} \sum_{E,v} \int \frac{d^4k}{(2 \pi)^4} \left[ - \frac{i}{E} (n^\chi_\perp \cdot \pa) \Phi_{E,v}^\chi (X,k) + \chi \frac{im}{2E^2} (\tilde{v} \cdot \pa) G^\chi_{E,v} (X,k)\right. \nn \\
& \left.  +\frac{i}{2E^2} \left(\tilde{v} \cdot k \ n^\chi_\perp \cdot \pa + k\cdot n^\chi_\perp \ \tilde{v} \cdot \pa - e \tilde{v}^\alpha F_{\alpha \mu} (X) n^\mu_{\perp,\chi} \right)  \Phi_{E,v}^\chi (X,k)\right]+ {\cal O} \left(\frac {1}{E^3} \right) \ , \label{eq:j5residual2}
\end{align}
where $\pa^\mu$ refers to the derivative in space $\pa_X^\mu$. These equations are equivalent to~(\ref{eq:sigmaresidual}) and (\ref{eq:j5residual}), but written in a convenient way for the later calculation of the axial anomaly.

At this stage it is possible to reconstruct the full momentum in Eqs.~(\ref{eq:sigmaresidual},\ref{eq:j5residual}) by using Eq.~(\ref{eq:Qpart}).  Note that the Wigner function includes a Dirac delta function that enforces the residual momentum to be on shell. Then,  $1/E_q$ is given by
\be \frac{1}{E_q} = \frac{1}{E+k\cdot u} \simeq \frac{1}{E} - \frac{\tilde{v}\cdot k}{2E^2} + {\cal O} \left( \frac{1}{E^3} \right) \ . \ee

One can further use~\cite{Manuel:2016wqs}
\be \sum_{E,v} \int \frac{d^4k}{(2\pi)^4}  = \int \frac{d^4q}{(2\pi)^4} \ , \ee
and introduce the vectors $n^\mu_{\perp,\chi,q}$ defined in Eq.~(\ref{eq:nperpq}).
We finally get
\begin{align}
\sigma(X) & =     \sum_{\chi=\pm} \int \frac{d^4q}{(2 \pi)^4}  \left[ \frac {2m}{E_q} G^\chi (X,q)- \chi \frac{i}{E_q} (n^\chi_{\perp,q} \cdot \pa) \Phi^\chi (X,q) \right. \nn \\
&- \left. \chi \frac{ie}{2E_q^2}   (2u^\alpha -  v_\chi^\alpha) F_{\alpha\mu}(X) n_{\perp,\chi,q}^{\mu} \Phi^\chi (X,q) \right] +  {\cal O} \left( \frac {1}{E_q^3} \right) \ , \label{eq:sigmafull} \\
j_5(X) & =   \sum_{\chi=\pm} \int \frac{d^4q}{(2 \pi)^4} \left[  -\frac{i}{E_q} (n^\chi_{\perp,q} \cdot \pa) \Phi^\chi (X,q)  + \chi \frac{im}{2E_q^2} (2 u - v_\chi) \cdot \pa G^\chi (X,q)\right. \nn \\
& \left. - \frac{ie}{2E_q^2}  (2u^\alpha-v_\chi^\alpha) F_{\alpha \mu}(X) n^\mu_{\perp,\chi,q} \Phi^\chi (X,q)\right]+ {\cal O} \left(\frac {1}{E_q^3} \right) \ , \label{eq:j5full}
\end{align}
where we have also substituted $\tilde{v}^\alpha$ by $2u^\alpha-v_\chi^\alpha$.

As we will see later, the scalar and pseudoscalar densities in the full theory are suppressed by a factor $1/E_q$ in comparison to the vector and axial functions.

It is also instructive (and a good check) to see that these functions remain invariant under RI transformations, as they are Lorentz scalars. 
In App.~\ref{app:typeIISigma} we focus on $\sigma(X)$ and its transformation under Type II up to ${\cal O}(1/E^3)$. We prove there that the scalar density is invariant under this transformation.

%%%%%%%%%%%%%%%%%%%%
\subsection{Vector and axial currents}
\label{macro-vector}
%%%%%%%%%%%%%%%%%%%%%%

We now detail the expression of the vector and the axial currents and perform a similar expansion in terms of the OSEFT fields. The currents are defined as
\begin{align}
j^\mu (x) &= \lim_{y \rightarrow x} \Tr \ \langle \bar \psi(y) \gamma^\mu \psi(x) \rangle  = \lim_{y \rightarrow x} \Tr \left\langle [ \bar {\chi}_v (y)+ \bar{H}^{(1)}_{\tilde v} (y) ] \gamma^\mu [   {\chi}_v (x)+ H^{(1)}_{\tilde v} (x) ]\right\rangle  \ , \\
j^\mu_5 (x) &= \lim_{y \rightarrow x} \Tr \ \langle \bar \psi(y)\gamma^\mu \gamma_5 \psi(x) \rangle  = \lim_{y \rightarrow x} \Tr \left\langle [ \bar {\chi}_v (y)+ \bar{H}^{(1)}_{\tilde v} (y) ]\gamma^\mu \gamma_5[   {\chi}_v (x)+ H^{(1)}_{\tilde v} (x) ]\right\rangle \ .
\end{align}

This particular calculation was already performed in Ref.~\cite{Carignano:2018gqt} for the massless case. Therefore we will not give full details here. The new terms depending on the fermion mass are   
\begin{align}
j_{\textrm{mass}}^\mu(X) & =   \sum_{\chi=\pm} \sum_{E,v} \int \frac{d^4k}{(2 \pi)^4}  \left[ -\frac {m^2 (v^\mu-\tilde{v}^\mu)}{2E^2} \ G_{E,v}^\chi (X,k) + \chi \frac{mi}{2E^2} (\tilde{v}\cdot \Delta_k) n^\mu_{\perp,\chi} \Phi_{E,v}^\chi (X,k) \right. \label{eq:jmures} \nn \\
&- \left. \chi \frac{mi\tilde{v}^\mu}{2E^2}   (n_{\perp}^\chi \cdot \Delta_k)  \Phi_{E,v}^\chi (X,k) \right] +  {\cal O} \left( \frac {1}{E^3} \right) \ ,\\
j_{5,\textrm{mass}}^\mu (X) & =  \sum_{\chi=\pm} \sum_{E,v} \int \frac{d^4k}{(2 \pi)^4} \left[  \frac{2m}{E} n^\mu_{\perp ,\chi} \Phi_{E,v}^\chi (X,k) - \frac{m}{E^2} \tilde{v}\cdot k \ n^\mu_{\perp ,\chi} \Phi_{E,v}^\chi (X,k) \right. \nn \\ 
& \left. - \chi \frac{m^2 (v^\mu+\tilde{v}^\mu )}{2E^2}  G^\chi_{E,v} (X,k) - \frac{m\tilde{v}^\mu}{E^2} (k \cdot n^\chi_\perp) \Phi_{E,v}^\chi (X,k)\right]+ {\cal O} \left(\frac {1}{E^3} \right) \ . \label{eq:jmu5res}
\end{align}

Adding these terms to the ones we already found in Ref.~\cite{Carignano:2018gqt} we can reconstruct the full variables. Notice that the on-shell velocity (\ref{onshellvector}) contains a term depending on the mass, which is precisely given by the first term in Eq.~(\ref{eq:jmures}), and part of the third one in Eq.~(\ref{eq:jmu5res}).

The final expressions for the vector and axial currents are 
\begin{align}
j^\mu(X) & =   \sum_{\chi=\pm} \int \frac{d^4q}{(2 \pi)^4}  \left\{ \left[ v_\chi^\mu - \frac{S_\chi^{\mu\nu} \Delta_\nu}{E_q}-\frac{e}{2} \frac{S_\chi^{\mu\nu}}{E_q^2} F_{\nu\rho}(X) (2u^\rho-v_\chi^\rho) \right] 2G^\chi(X,q) \right. \nn \\
& \left. + \chi \frac{mi}{E^2_q} (u \cdot \Delta) n^\mu_{\perp,\chi,q} \Phi^\chi (X,q) 
- \chi \frac{mi(2u^\mu -v_\chi^\mu)}{2E_q^2}   (n_{\perp,q}^\chi \cdot \Delta)  \Phi^\chi (X,q) \right\} +  {\cal O} \left( \frac {1}{E_q^3} \right) \ , \label{eq:j} \\
j_{5}^\mu (X) & =  \sum_{\chi=\pm} \int \frac{d^4q}{(2 \pi)^4} \left\{ \chi \left[ v_\chi^\mu - \frac{S_\chi^{\mu\nu} \Delta_\nu}{E_q}-\frac{e}{2} \frac{S_\chi^{\mu\nu}}{E_q^2} F_{\nu\rho} (X) (2u^\rho-v_\chi^\rho) \right] 2G^\chi(X,q) \right. \nn \\
& \left. + \frac{2m}{E_q} n^\mu_{\perp ,\chi,q} \Phi^\chi (X,q) 
- \chi \frac{m^2 (2u^\mu-v_\chi^\mu)}{E_q^2}  G^\chi (X,q)  \right\} + {\cal O} \left(\frac {1}{E_q^3} \right) \ , \label{eq:j5}
\end{align}
where $v_\chi^\mu$ is given by Eq.~(\ref{onshellvector}). We stress again that, as opposed to the (pseudo)scalar densities, the currents do contain a term ${\cal O} (1/E_q^0)$. The reason is that while $\bar{\chi}_v(y) \chi_v(x)$ identically vanishes, $\bar{\chi}_v(y) \gamma^\mu \chi_v(x)$ does not.

%%%%%%%%%%%%%%%%%%%%%%
\subsection{Correspondence with previous approaches}
\label{dirac-rep}
%%%%%%%%%%%%%%%%%%%%%%

  Several works have considered the effects of the fermion mass in the kinetic theory, using the Dirac Lagrangian as the starting point~\cite{Zhuang:1995pd,Hattori:2019ahi,Gao:2019znl,Weickgenannt:2019dks,Wang:2019moi,Li:2019qkf,Gao:2019zhk,Weickgenannt:2020sit,Yang:2020hri,Liu:2020flb}. While their formalism deviates from ours (we have extensively described the differences between the Pauli-Dirac and the OSEFT representations in~\cite{Carignano:2019zsh}), the generic structure of the Wigner function should have a correspondence with the OSEFT formalism (as both fermion degrees of freedom are related, cf. Eq.~(\ref{eq:Fields})). In particular, we want to study the equivalence of the quantum coherence function $\Phi^\chi (X,q)$ in the representation used by these works.
  
  After a look to the formalism in~\cite{Zhuang:1995pd, Wang:2019moi} it is clear that the main effect of the finite fermion mass enters through the functions ${\bf g}_2 (X,{\bf q})$ and ${\bf g}_3 (X,{\bf q})$. These are functions of 4-position and 3-momentum as they are defined via the equal-time Wigner function (see~\cite{Zhuang:1995pd, Wang:2019moi} for details). Nevertheless, one can always express these functions in terms of the standard (covariant) Wigner function,
\be W(X,q)=\int d^4s \ e^{i q\cdot s} \ \left\langle \bar{\psi} \left( X+ \frac{s}{2} \right) U \left( X+ \frac{s}{2},X-\frac{s}{2} \right) \psi \left( X- \frac{s}{2} \right) \right\rangle \ . \ee
The following expressions are taken from Ref.~\cite{Zhuang:1995pd},
\begin{align}
g^i_0 (X,{\bf q})  & =-\int dq^0 A^i(X,q) = -\int dq^0 \ \textrm{Tr} (\gamma^i \gamma^5 W(X,q)) \\
g^i_2 (X,{\bf q}) & = -\int dq^0 S^{0i} (X,q) = -\int dq^0 \ \textrm{Tr} (\sigma^{0i} W(X,q)) \\
g^i_3 (X,{\bf q}) & = \frac12 \epsilon^{ijk} \int dq^0 S^{jk} (X,q) =
\frac12 \epsilon^{ijk} \int dq^0 \ \textrm{Tr} (\sigma^{jk} W(X,q))
\end{align}
where the functions $A^\mu,S^{\mu\nu}$ are defined in Ref.~\cite{Zhuang:1995pd} and $\sigma^{\mu\nu}$ is defined after Eq.~(\ref{Dirac-decomposition}). Now $W(X,q)$ should be expressed in terms of the OSEFT Wigner function $S^<_{E,v}(X,q)$ using the $1/E$ expansion. On one hand it is straightforward to make the connection between ${\bf g}_0 (X,{\bf q})$ and $j_5^i(X,q)$ given in Eq.~(\ref{eq:j5}),
\begin{align}
g^i_0 (X,{\bf q})  &  = -\int dq^0 j_5^i (X,q) \ , \label{eq:g0}
\end{align}
where the OSEFT fields are written in terms of the full momentum. On the other hand we also find
\begin{align}
g^i_2 (X,{\bf q}) & =2i \int dq^0  \sum_{\chi=\pm} \chi (v \wedge n_{\perp,\chi,q})^{0i} \Phi^\chi (X,q) + {\cal O} \left( \frac{1}{E} \right) \ , \\
g^i_3 (X,{\bf q}) & =-i \epsilon^{ijk} \int dq^0  \sum_{\chi=\pm} \chi (v \wedge n_{\perp,\chi,q})^{jk} \Phi^\chi (X,q) + {\cal O} \left( \frac{1}{E} \right) \ ,
\end{align}
at lowest order in the $1/E$ expansion. We have used Eqs.~(\ref{eq:tracesigma1},\ref{eq:tracesigma2}). Therefore, the functions ${\bf g}_2(X,{\bf q})$ and ${\bf g}_3 (X,{\bf q})$ are indeed directly related to the spin coherence function.

These expressions can be further simplified in the LRF, there the wedge product can be directly written in terms of the vectors $n^i_{\perp,\pm,q}$. Using the conventions described before Eq.~(\ref{eq:GT}) we obtain
\begin{align}
g^i_2 (X,{\bf q}) & =2i \int dq^0  \sum_{\chi=\pm} \chi n_{\perp,\chi,q}^i \Phi^\chi (X,q) + {\cal O} \left( \frac{1}{E} \right) \ ,  \label{eq:g2} \\
g^i_3 (X,{\bf q}) & =-2 \int dq^0  \sum_{\chi=\pm} n_{\perp,\chi,q}^i \Phi^\chi (X,q) + {\cal O} \left( \frac{1}{E} \right) \ ,  \label{eq:g3}
\end{align}
or equivalently (using ${\bf n}_{\perp,+,q} \cdot {\bf n}_{\perp,-,q}=2  $ at leading order)
\begin{align}
\int dq^0 \ \Phi (X,q) & = \frac{i}{8} {\bf n}_{\perp,+,q} \cdot [ {\bf g}_2 (X,{\bf q}) + i {\bf g}_3 (X,{\bf q}) ] + {\cal O} \left( \frac{1}{E} \right) \ ,  \\
\int dq^0 \ \Phi^\dag (X,q) & = -\frac{i}{8} {\bf n}_{\perp,-,q} \cdot [ {\bf g}_2 (X,{\bf q}) - i {\bf g}_3 (X,{\bf q}) ] + {\cal O} \left( \frac{1}{E} \right) \ .  
\end{align}

As an additional nontrivial check, we can show that the lowest-order equations of motion in~\cite{Zhuang:1995pd,Wang:2019moi} are indeed consistent with ours. In the classical limit [at ${\cal O} (\hbar^0)$] the functions ${\bf g}_2(X,{\bf q})$ and ${\bf g}_3(X,{\bf q})$ are related to ${\bf g}_0(X,{\bf q})$~\cite{Zhuang:1995pd},
\begin{align}
{\bf g}_2^{+ (0)} (X,{\bf q}) &= \frac{{\bf q} \times {\bf g}_0^{+ (0)} (X,{\bf q})}{m} \ , \label{eq:g2eom}\\
{\bf g}_3^{+ (0)} (X,{\bf q}) &=  \frac{ E^2_q {\bf g}_0^{+ (0)} (X,{\bf q}) - E_q [ {\bf q} \cdot {\bf g}_0^{+ (0)} (X,{\bf q})] {\bf q}}{mE_q } \ , \label{eq:g3eom}
\end{align}
where we have already particularized for their positive-energy states (consistent with our particle sector). At leading order in $1/E$ one can simply replace ${\bf q}=E_q {\bf v}_q$.

From Eq.~(\ref{eq:g0}), this indicates that both ${\bf g}_2(X,{\bf q})$ and ${\bf g}_3(X,{\bf q})$ can be written in terms of the axial current (\ref{eq:j5}). In the classical limit, the only term from Eq.~(\ref{eq:j5}) that contributes to these functions is the one depending on the spin coherence function. Inserting this term into~(\ref{eq:g2eom}) and~(\ref{eq:g3eom}), and particularizing to the LRF---where ${\bf v}_q \times {\bf n}_{\perp,\chi,q} = - i\chi {\bf n}_{\perp,\chi,q}$ and ${\bf v}_q \cdot {\bf  n}_{\perp,\chi,q}=0$ at leading order---these equations precisely give Eqs.~(\ref{eq:g2}) and (\ref{eq:g3}), providing a consistency check of the equations of motion at lowest order/classical limit.

It would be interesting to make a profound comparison study between the results of the OSEFT, and the Dirac formalism used in Refs.~\cite{Zhuang:1995pd, Wang:2019moi}. However, such an analysis goes beyond the scope of this work.

%%%%%%%%%%%%%%%%%%%%%%%%%%%
\section{Chiral anomaly in the presence of a small mass term}
\label{chiral-anomaly}
%%%%%%%%%%%%%%%%%%%%%%%%%%

In this section we will consider the 4-divergence of the vector and axial currents to check that: 1) the vector current is conserved, and 2) we reproduce the axial anomaly expression with fermion  mass corrections.

 Let us first start with the vector current by taking the divergence of the current in Eq.~(\ref{eq:j}). In terms of the full variables this divergence reads,
\begin{align}
 \pa_\mu j^\mu (X)  & =  \sum_{\chi=\pm}    \int \frac{d^4q}{(2 \pi)^4} \left[ v_\chi \cdot \pa + \frac{1}{E_q} S_\chi^{\mu\nu} \pa_\mu \Delta_\nu - \frac{e}{2E_q^2} S_\chi^{\mu\nu} \pa_\mu F_{\nu\rho} (X) (2u^\rho - v_\chi^\rho) \right]  2 G^\chi(X,q)  \nn \\
 & +   \sum_{\chi=\pm} \int \frac{d^4q}{(2 \pi)^4} \left[
 \chi \frac{mi}{2E^2_q} u_\nu \pa_\mu \Delta^\nu n^\mu_{\perp,\chi,q} \Phi^\chi (X,q) 
- \chi \frac{mi (2u^\mu-v_\chi^\mu)}{2E_q^2}   n_{\perp,\nu,q}^\chi \pa_\mu \Delta^\nu  \Phi^\chi (X,q) \right] \nn \\
& + {\cal O} \left(\frac {1}{E_q^3} \right) \ , 
\end{align}
where the first line coincides with the result of Ref.~\cite{Carignano:2018gqt} for the massless case, with the difference that $v_\chi^\mu$ now contains a mass dependent term, as shown in Eq.~(\ref{onshellvector}). To see the equivalence, one has to make use of the gradient expansion and neglect the second term ${\cal O}(\pa \Delta)$ in the first line, because it is ${\cal O}(\pa_X^2)$ or ${\cal O} (\pa X\pa_q)$, and therefore suppressed. This also means that the terms proportional to $m$ in the second line should not be considered. Overall the divergence of the vector current reduces to the same expression of Ref.~\cite{Carignano:2018gqt},
\be \pa_\mu j^\mu (X)=0 \ . \ee

Let us now turn to the case of the axial current divergence, and see that some nontrivial terms are generated. 
We start from Eq.~(\ref{eq:j5})  We apply $\pa_\mu$ to the axial current and, as for the vector current, we already neglect terms subleading in the gradient expansion, i.e. ${\cal O}(\pa \Delta)$. We obtain, using the transport equation for the vectorial components,
\begin{align}
 \pa_\mu j_5^\mu (X) &  =   \sum_{\chi=\pm} \chi \int \frac{d^4q}{(2 \pi)^4} \left[ ev_\chi^\mu F_{\mu\nu} (X)\pa_q^\nu  - \frac{e^2}{2E_q^2} S_\chi^{\mu\rho}  F_{\rho\sigma} (X) (2u^\sigma - v_\chi^\sigma) F_{\mu\nu} (X) \pa_q^\nu \right]  2 G^\chi(X,q) \nn \\
 & + 2mij_5(X) \nn \\
 &+ i \sum_{\chi=\pm} \int \frac{d^4q}{(2 \pi)^4}  \frac{e}{2E^2_q} (2u^\alpha -v^\alpha_\chi) F_{\alpha\mu} (X) n^\mu_{\perp,\chi} \Phi^\chi (X,q) + {\cal O} \left(\frac {1}{E_q^3} \right) \ , \label{eq:divj5}
\end{align}

The first line of Eq.~(\ref{eq:divj5}) reproduces the same steps in Ref.~\cite{Carignano:2018gqt}, but here with mass corrections included in $v_\chi^\mu$. For a thermal plasma, with Fermi-Dirac  distribution functions, one can evaluate the momentum integrals. We remind that to obtain it, it is easier to work in the LRF and integrate over $q_0$ using the on-shell condition incorporated in $G^\chi (X,q)$. Then, one can integrate by parts in $d^3q$ to get the surface contribution of the second term (the one quadratic in $F_{\mu\nu}$). The addition of both particle and antiparticle contributions in the remaining integration over $dq$ with a thermal distribution for $f^\chi(X,q)$ reproduces the consistent version of the chiral anomaly. The last term in (\ref{eq:divj5}) gives zero when doing the same steps because of the angular integration.

The divergence of the axial current, including the contribution of both particles and antiparticles, depends on the IR limit of the thermal distribution function~\cite{Son:2012wh,Stephanov:2012ki,Manuel:2014dza,Gao:2019zhk}
\be 
\pa_\mu {\cal J}_5^\mu (X)  = \frac16 \frac{e^2}{2\pi^2} {\bf E} (X) \cdot {\bf B} (X) \sum_{\chi=\pm} \lim_{|{\bf q}| \rightarrow 0}  \ [f^\chi(E^\chi_q) + \tilde{f}^\chi ( E^\chi_q)] \ +2mi {\cal J}_5(X) \ ,
\ee
where ${\cal J}_5^\mu$ and ${\cal J}_5$ are the total (particle plus antiparticle contributions) axial current and pseudoscalar densities, respectively.

While a generic (non equilibrium) result for the axial anomaly is unfortunately not possible---as it depends on the population of very soft fermion modes---we can check the equilibrium case by applying the limit to the Fermi-Dirac distribution function
\be 
f^\chi( E^\chi_q ) =\frac{1}{e^{( E^\chi_q -\mu_\chi)/T} +1} \ , \qquad   {\tilde f}^\chi( E^\chi_q ) =\frac{1}{e^{( E^\chi_q +\mu_\chi)/T} +1} \ ,
\ee
where $E^\chi_q=|{\bf q}|$ at lowest order, or the one given by the dispersion relation Eq.~(\ref{eq:onshell}). In both cases one finally arrives to~\cite{Manuel:2014dza,Carignano:2018gqt,Gao:2019zhk},
\be
 \pa_\mu {\cal J}_5^\mu (X)  = \frac13 \frac{e^2}{2\pi^2} {\bf E} (X) \cdot {\bf B} (X) +2mi {\cal J}_5(X) \ .
\ee
 We should stress that we have ignored in this section the presence of axial gauge fields.

%%%%%%%%%%%%%%%%
\section{Outlook}
\label{Conclu}
%%%%%%%%%%%%%%%%%

In this work we have studied how the chiral kinetic theory, as derived from the OSEFT, is modified by small mass corrections. While in the massless case only 
four  Wigner functions are needed, corresponding to  both particles and antiparticles  of the two possible fermion chiralities, as soon as a small mass of the fermions is considered, four extra Wigner functions are required for the complete treatment of the system. 
We have checked the consistency of the whole approach, by showing that it respects RI, and ultimately, the Lorentz symmetry, and that it allows for describing the small mass modifications to the chiral anomaly. We also showed the correspondence of the spin coherence functions to analogous terms found in previous works, showing the consistency of the equations of motion at the classical level.

Our transport equations can be considered as a generalization of the neutrino transport equations of Ref.~\cite{arXiv:1605.09383}. In those works neutrinos were assumed  only to contact interact with electrons and baryons, while  gauge interactions were not considered in the energy regime considered. Our work fills that gap, that might be needed for the consideration of neutrinos in
cosmological settings. However, our transport approach might be useful to describe other systems with almost chiral fermions, such as the quark-gluon plasma~\cite{Huang:2015oca}, or quasiparticles in Weyl semimetals~\cite{Miransky:2015ava}.

It can be particular interesting to see how the dynamical evolution of a chiral plasma is affected by the presence of a small mass, and how it might affect, for example, the chiral plasma instabilities~\cite{Akamatsu:2013pjd,Tuchin:2014iua,Manuel:2015zpa}. A mass term typically induces a  damping term in the chiral anomaly equation, and this is an effect that has been considered in several publications~\cite{Giovannini:1997eg,Boyarsky:2011uy,Manuel:2015zpa,Hou:2017szz,Boyarsky:2020cyk,Boyarsky:2020ani}. Our work reveals that it is also necessary to introduce the effect of the quantum coherent states, which might modify the dynamical evolution of the chiral density at shorter time scales. We leave these studies for future projects.

\begin{acknowledgments}

We have been supported by the Ministerio de Ciencia  e Innovaci\'on (Spain) under the projects FPA2016-81114-P and  PID2019-110165GB-I00,  as well as by the project  2017-SGR-929  (Catalonia), the Deutsche  Forschungsgemeinschaft  (DFG,  German  Research  Foundation) through Projects No. 411563442 (Hot Heavy Mesons)  and  No.  315477589  -  TRR  211  (Strong-interaction matter under extreme conditions). This work was also supported by the COST Action CA15213 THOR.
 
\end{acknowledgments}

\begin{appendix}

\section{Wigner transform of fields and gradient expansion~\label{app:Wigner}}

For any two-point function like $G_{E,v}^\chi(x,y),\Phi_{E,v}^\chi(x,y),\rho_G^\chi(x,y),\rho_\Phi^\chi(x,y)$ one can define the (gauge invariant) Wigner transform (WT) as
\be
G^\chi_{E,v} (x,y) \WT G_{E,v}^\chi (X,\tilde{k}) \equiv \int d^4s e^{i k\cdot s}  U(x,y)  G^\chi_{E,v}(x,y) \ , \ee
where the resulting field is a function of the center-of-mass coordinate $X=(x+y)/2$, and the relative kinetic momentum $\tilde{k}^\mu = k^\mu -eA^\mu (X)$, after Fourier transform of the relative coordinate $s=x-y$. The gauge link (Wilson line) $U(x,y)$ helps to define a gauge-invariant transformed quantity. In what follows (and everywhere in the main text) we have denoted the kinetic momentum simply as $k$, suppressing the tilde without risk of confusion.

When differential operators act on two-point functions they should be transformed as well. For example, the action of the covariant derivative operator $iD^\mu_x=i\pa^\mu_x-eA^\mu(x)$ over $G^\chi_{E,v}(x,y)$ has the following Wigner transform,
\begin{align} 
  iD^\mu_{x} G^\chi_{E,v} (x,y) & \WT \int d^4s e^{ik\cdot s} [i\pa^\mu_x-eA^\mu (x)] U(x,y) G_{E,v}^\chi (x,y) \\
 & =  \int d^4s e^{ik\cdot s} \left[ \frac{i}{2} \pa^\mu_X +i\pa^\mu_s-eA^\mu (x) \right] \ e^{-i e s\cdot A(X+s/2)} \ G_{E,v}^\chi \left( X+\frac{s}{2},X-\frac{s}{2} \right) \ ,
\end{align}
where we have implemented a gauge link of the form $U(x,y) = \exp[- ie (x-y) \cdot A(x)]$. 

To simplify the final expressions we perform a gradient expansion in $\pa^\mu_X$ and keep the lowest order terms, for example
\be A^\mu(x) =A^\mu \left( X+\frac{s}{2} \right)= A^\mu(X) + \frac{1}{2} s_\nu \pa^\nu_X A^\mu(X) + \cdots  \ee

In this situation one can write
\begin{align} 
  iD^\mu_{x} G^\chi_{E,v} (x,y) & \WT \int d^4s e^{i s\cdot (k - e A(X))} \left[ \frac{i}{2} \pa^\mu_X+i\pa^\mu_s + \frac{e}{2} s_\nu F^{\mu\nu}(X) \right] G_{E,v}^\chi \left(X+\frac{s}{2},X-\frac{s}{2} \right) \nn \\ 
  & = \left( \frac{i}{2} \Delta_k^\mu + k^\mu \right) G^\chi_{E,v}(X,k) \ ,
\end{align}
where $\Delta_k^\mu=\pa_X^\mu-eF^{\mu\nu}(X)\pa_{k,\nu}$.

In practical calculations we will usually find a combination of operators acting on both $x$ and $y$, e.g. $iD_x^\mu \mp (iD_y^\mu)^*$. The Wigner transforms of such combinations are
\begin{align}
 [ iD_x^\mu - (iD_y^\mu)^* ] G_{E,v}^\chi (x,y) & \WT  i\Delta_k^\mu G_{E,v}^\chi (X,k) \ , \label{eq:op1minus} \\
 [ iD_x^\mu + (iD_y^\mu)^* ] \rho_G^\chi (x,y) & \WT  2k^\mu  \rho_G^\chi (X,k) \ . \label{eq:op1plus} 
\end{align}

More involved operators are formally treated the same way. For the calculation of the chiral kinetic equation we find the following Wigner transforms,
\begin{align}
[ iD_x^\mu iD_x^\nu - (iD^\mu_y)^* (iD_y)^*  ]G_{E,v}^\chi (x,y) & \WT i[k^\mu \Delta_k^\nu+k^\nu \Delta_k^\mu -e F^{\mu\nu} (X)] G^\chi_{E,v}(X,k) \ , \label{eq:op2minus} \\
[ i D^\alpha_x D^2_{x} - (i D^\alpha_y D^2_y)^* ]G_{E,v}^\chi (x,y) & \WT [k^2\Delta_k^\alpha+2k^\alpha k\cdot \Delta_k +2iek_\mu F^{\alpha\mu} (X)] G^\chi_{E,v}(X,k) \ . \label{eq:op3minus} 
\end{align}

For the calculation of the dispersion relations one needs to apply
\begin{align}
[ iD_x^\mu iD_x^\nu + (iD^\mu_y)^* (iD_y)^* ] \rho_G^\chi (x,y) & \WT 2 k^\mu k^\nu  \rho_G^\chi(X,k) \ , \label{eq:op2plus} \\
[ i D^\alpha_x D^2_{x} + (i D^\alpha_y D^2_y)^* ] \rho_G^\chi (x,y) &  \WT [-eF^{\alpha \mu} (X) \Delta_{k,\mu}-2k^\alpha k^2] \rho_G^\chi(X,k) \ . \label{eq:op3plus} 
\end{align}

Relying on the gradient expansion we have neglected terms ${\cal O}(\pa_X^2)$ and ${\cal O}(\pa_X \pa_k)$, like $\pa_X^\mu F_{\mu\nu}(X)$, for instance.

%%%%%%%%%%%%%%%%%%%%%%%%%%%
\section{Dirac traces~\label{app:traces}}
%%%%%%%%%%%%%%%%%%%%%%%%%%

We  provide here some Dirac traces needed in several parts of this work. Some of them were already used for the vectorial components in the massless case in our previous paper~\cite{Carignano:2018gqt}. The Dirac traces needed for such components, but in the massive case read
\begin{align}
  \Tr \left[\frac{\slashed{\tilde v} }{2} P_\chi S^<_{E,v} (x,y)\right] &= 2 G_{E,v}^\chi (x,y) \ , \label{trace-G1} \\
  \Tr \left[\sigma_\perp^{\mu\nu} \frac{\slashed{\tilde v} }{2} P_\chi S^<_{E,v} (x,y)\right] &=  \chi \epsilon^{\mu_\perp \nu_\perp \alpha  \beta }  v_\alpha \tilde{v}_\beta G_{E,v}^\chi (x,y) \ ,  \label{trace-G2} \\
   \Tr \left[\gamma^\mu_\perp \frac{\slashed{\tilde v} }{2} P_\chi S^<_{E,v} (x,y)\right] &= -2 \chi  n_{\perp,\chi}^\mu \Phi_{E,v}^\chi (x,y) \ ,  \label{trace-G3}
 \end{align} 
where the last one was absent in the massless case~\cite{Carignano:2018gqt}. When applied to the calculation of the dispersion relation of the vector/axial components, $S^<_{E,v} (x,y)$ is to be replaced by the spectral function $\rho(x,y)$, and $G^\chi_{E,v}(x,y)$ by $\rho^\chi_G(x,y)$ [and $\Phi^\chi_{E,v}(x,y)$ by $\rho_\Phi^\chi(x,y)$].

For the derivation of the kinetic equation obeyed by the tensorial components of the Wigner function one needs to compute similar Dirac traces. These can be simplified using the relation
\be
\slashed{n}_{\perp,\chi}\slashed{\tilde v}= i \sigma_{\alpha \beta}\, {\tilde v}^\alpha n_{\perp,\chi}^\beta \ .
\ee
In the Wigner function $S^<_{E,v}(x,y)$ one has two chiral tensorial components, which are immediately selected by projecting with $n_{\perp, \chi}$. The vector $n_{\perp,+}$ selects the $L$ component (remember $n_{\perp,+} \cdot  n_{\perp,+}  = 0$). The needed traces for this sector are,
 \begin{align}
 \Tr \left[\slashed{n}_{\perp,\chi}\frac{\slashed{\tilde v} }{2} S^<_{E,v} (x,y)\right] &= - 4 \chi\Phi_{E,v}^{-\chi} (x,y) \ , \label{eq:trace1-Phi1} \\
\Tr \left[\slashed{n}_{\perp,\chi} \gamma^\mu_\perp \frac{\slashed{\tilde v} }{2} S^<_{E,v} (x,y)\right] &=  \sum_{\tilde \chi=\pm} \left( 2n_{\perp,\chi}^\mu +i{\tilde \chi} \epsilon^{\mu_\perp \nu \alpha \beta  } v_\alpha \tilde{v}_\beta  n^\perp_{\nu,\chi} \right) G_{E,v}^{\tilde \chi}(x,y) =4 n_{\perp,\chi}^\mu G_{E,v}^{\chi} (x,y)\ , \label{eq:trace1-Phi2} \\
\Tr \left[ \slashed{n}_{\perp,\chi} \sigma^{\mu\nu}_\perp \frac{\slashed{\tilde v}}{2} S^<_{E,v}(x,y) \right]  &=  2 i 
(n_{\perp,+} \wedge n_{\perp,-})^{\mu \nu}  \Phi_{E,v}^{-\chi} (x,y) \ ,  \label{eq:trace1-Phi3}
\end{align}
where to simplify the trace (\ref{eq:trace1-Phi2}) we have used the useful relation
\be \label{eq:trick}
 i  \epsilon^{\mu_\perp \nu \alpha \beta} v_\alpha  n^{\perp,\chi}_{\nu} = -  ((v \wedge n_{\perp,\chi})^{\mu_\perp \beta})^\star =  \chi v^\beta n_{\perp,\chi}^{\mu} \ .
 \ee 
Similar traces with a $\gamma_5$ insertion are also required,
\begin{align}
\Tr \left[\slashed{n}_{\perp,\chi}\frac{\slashed{\tilde v} }{2} \gamma_5 S^<_{E,v} (x,y)\right] &= 4 \Phi_{E,v}^{-\chi} (x,y) \ , \label{eq:traceg5-Phi1} \\
\Tr \left[\slashed{n}_{\perp,\chi} \gamma^\mu_\perp \frac{\slashed{\tilde v} }{2} \gamma_5 S^<_{E,v} (x,y)\right] &=4\chi n_{\perp,\chi}^\mu G_{E,v}^{\chi}(x,y) \ , \label{eq:traceg5-Phi2} \\
\Tr \left[ \slashed{n}_{\perp,\chi} \sigma^{\mu\nu}_\perp \frac{\slashed{\tilde v}}{2} \gamma_5 S^<_{E,v}(x,y) \right]  &=  -2 i \chi (n_{\perp,+} \wedge n_{\perp,-})^{\mu \nu}  \Phi_{E,v}^{-\chi} (x,y) \ . \label{eq:traceg5-Phi3}
\end{align}
When considering the dispersion relation of the tensor component of the Wigner function one needs to replace $S^<_{E,v}(x,y)$ by $\rho(x,y)$, and $\Phi_{E,v}^\chi(x,y)$ by $\rho_\Phi^\chi(x,y)$ [and $G^\chi_{E,v} (x,y)$ by $\rho^\chi(x,y)$].

\section{Components of the kinetic equation and dispersion relation~\label{app:details}}

We summarize here the different functions used to determine the components of the kinetic equations and dispersion relations, for both $G_{E,v}^\chi(x,y)$ and $\Phi^\chi_{E,v}(x,y)$. We do not detail the full derivation because it is very similar to the one already sketched in Ref.~\cite{Carignano:2018gqt}  However, an analogous step-by-step calculation is given in the App.~\ref{app:scalarden} for the scalar fermion density $\sigma(X)$. 
 
We begin with the functions $I_{-}^{(n),\chi}$ needed to construct the kinetic equation of $G^\chi_{E,v}(X,k)$, which are defined in Eq.~(\ref{eq:Iminus}). Up to $n=2$, and in terms of the residual momentum, they read
\begin{align}
 I_{-}^{(0),\chi} &= 2 i v^\mu \Delta_{k,\mu} G^\chi_{E,v}(X,k)  \ , \\
 I_{-}^{(1),\chi} & =  \frac{2i}{E}  {k}_{\perp}^\mu \Delta_{k,\mu} G^\chi_{E,v}(X,k) \ , \\
 I_{-}^{(2),\chi} &= 
 \frac{2i}{E^2} \left\{ - k^\mu_{\perp} \frac{{\tilde v}\cdot k - v \cdot k}{2}   + \frac 14 \left[  k^2_{\perp} -m^2- \frac {e\chi}{4}  \epsilon^{\alpha \beta \delta \gamma} {\tilde v}_\beta v_{\alpha} F^\perp_{\delta \gamma} \right] (v^\mu - {\tilde v}^\mu ) \right. \nn \\
 & \left. - \frac{e \chi}{8} \epsilon^{\alpha \beta \mu_\perp \nu_\perp} {\tilde v}_\beta v_{\alpha} F_{\nu \rho} {\tilde v}^\rho  \right\} \Delta_{k,\mu} G^\chi_{E,v}(X,k) 
\ ,
\end{align}
where for $n=0$ one needs to use the Dirac trace~(\ref{trace-G1}) and the Wigner transform~(\ref{eq:op1minus}); for $n=1$ one should apply the same trace~(\ref{trace-G1}) and the Wigner transform~(\ref{eq:op2minus}) (with the appropriate contractions); and for $n=2$ one uses~(\ref{trace-G1},\ref{trace-G2}) as well as~(\ref{eq:op1minus}) and (\ref{eq:op3minus}).

For the dispersion relation satisfied by the vector and axial vector components of the Wigner functions one needs the functions $I_{+}^{(n),\chi}$, defined in Eq.~(\ref{eq:Iplus}). The result for these are 
\begin{align}
I_{+}^{(0),\chi} & = 4 k \cdot v  \,  \rho_G^\chi (X,k) \ , \\
I_{+}^{(1),\chi} & =  \frac{2}{E}  \left[  k^2_{\perp} -m^2  - \frac{e \chi}{4} \epsilon^{\alpha \beta \mu_\perp \nu_\perp} {\tilde v}_\beta v_{\alpha} F_{\mu \nu} (X)
  \right] \rho_G^\chi (X,k) \ , \\
I_{+}^{(2),\chi} & =  
-\frac{2}{E^2} \left\{ \left[  k^2_{\perp}-m^2  - \frac{e \chi}{4} \epsilon^{\alpha \beta \mu_\perp \nu_\perp} {\tilde v}_\beta v_{\alpha} F_{\mu \nu} (X) \right]  \frac{{\tilde v}\cdot k - v \cdot k}{2} \right. \nn \\
&\left.  + \frac{e \chi}{4}\epsilon^{\alpha \beta \mu_\perp \nu_\perp} {\tilde v}_\beta v_{\alpha} F_{\nu \rho}(X) {\tilde v}^\rho k_{\mu}   \right\}  \rho_G^\chi (X,k) \nonumber + i\frac{em \chi}{E^2}  {\tilde v}^\mu F_{\mu \alpha} (X) n^\alpha_{\perp, \chi} \ \rho_\Phi^{\chi}(X,k) \ ,
\end{align}
where for $n=0$ one needs to use the Dirac trace~(\ref{trace-G1}) and the Wigner transform~(\ref{eq:op1plus}); for $n=1$ one should apply the traces (\ref{trace-G1},\ref{trace-G2}) and the Wigner transform~(\ref{eq:op2plus}); and for $n=2$ one uses the 3 traces~(\ref{trace-G1},\ref{trace-G2},\ref{trace-G3}) as well as Dirac traces~(\ref{eq:op1plus}) and (\ref{eq:op3plus}).

We now turn to the tensorial components of the Wigner function. To construct their kinetic equation one requires the functions $I^{(n)}_{\Phi^\chi;-}$ defined in Eq.~(\ref{eq:IPhiminus}) (recall that $\Phi_{E,v}^-=\Phi_{E,v}, \Phi_{E,v}^+=\Phi_{E,v}^\dag$). Up to $n=2$ they are
\begin{align}
I_{\Phi^\chi,-}^{(0)} &=  4 i \chi \ v \cdot \Delta_k \ \Phi_{E,v}^\chi (X,k)  \ , \\
I_{\Phi^\chi,-}^{(1)} & =  \frac{4i\chi}{E} \ {k}_{\perp} \cdot  \Delta_k \ \Phi_{E,v}^\chi (X,k) \ , \\
I_{\Phi^\chi,-}^{(2)} &= 
 \frac{4i\chi}{E^2} \left\{ - k^\mu_{\perp} \frac{{\tilde v}\cdot k - v \cdot k}{2}   + \frac 14 \left[  k^2_{\perp} -m^2 - \frac {ie\chi}{4}  (n_{\perp,+} \wedge n_{\perp,-})^{\alpha\beta} F_{\alpha \beta} (X) \right] (v^\mu-\tilde{v}^\mu)  \right. \nn \\
 & \left.  - \frac{ie\chi}{8} (n_{\perp,+} \wedge n_{\perp,-})^{\mu\nu} F_{\nu\rho} (X)\tilde{v}^\rho \right\} \Delta_{k,\mu} \Phi_{E,v}^\chi(X,k)]   \ ,
\end{align}
where for $n=0$ one needs to use the Dirac traces~(\ref{eq:trace1-Phi1},\ref{eq:traceg5-Phi1}) and the Wigner transform~(\ref{eq:op1minus}); for $n=1$ one should apply traces (\ref{eq:trace1-Phi2},\ref{eq:traceg5-Phi2}) and the Wigner transform~(\ref{eq:op2minus}); and for $n=2$ one uses the traces~(\ref{eq:trace1-Phi1},\ref{eq:traceg5-Phi1},\ref{eq:trace1-Phi2},\ref{eq:traceg5-Phi2},\ref{eq:trace1-Phi3},\ref{eq:traceg5-Phi3}) and the Wigner transforms (\ref{eq:op1minus},\ref{eq:op3minus}). Of course, in all formulas of the Wigner transform one needs to replace $G_{E,v}^\chi$ by $\Phi_{E,v}^\chi$.

For the dispersion relation of the tensorial components one requires to know the functions $I_{\Phi^\chi;+}^{(n)}$ defined in Eq.~(\ref{eq:IPhiplus}). Up to $n=2$ they read,
\begin{align}
I_{\Phi^\chi;+}^{(0)} & = 8\chi \ k \cdot v  \ \rho_\Phi^\chi (X,k) \ ,  \\
I_{\Phi^\chi;+}^{(1)} & =  \frac{4\chi}{E}  \left[  k^2_{\perp} -m^2  - i \chi \frac{e}{4}  (n_{\perp,+} \wedge n_{\perp,-} )^{\mu\nu} F_{\mu\nu} (X) \right] \rho_\Phi^\chi (X,k) \ ,   \\
I_{\Phi^\chi;+}^{(2)} & =  
-\frac{4\chi}{E^2} \left\{ \left[  k^2_{\perp}-m^2  - i\chi \frac{e}{4}  (n_{\perp,+} \wedge n_{\perp,-})^{\mu\nu} F_{\mu\nu} (X) \right] \frac{\tilde{v} \cdot k-v\cdot k}{2} \right. \nn \\
& \left. + i\chi \frac{e}{4} (n_{\perp,+} \wedge n_{\perp,-})^{\mu\nu} F_{\nu\rho}(X) \tilde{v}^\rho k_\mu  \right\} \rho_\Phi^\chi(X,k) - i\frac{2me}{E^2} \tilde{v}^\mu F_{\mu\alpha}(X)  n_{\perp,-\chi}^\alpha \ \rho_G^{-\chi} (X,k)  \ , \label{eq:IPhi2plus}
\end{align}
where we have used Eq.~(\ref{eq:trick}) to simplify (\ref{eq:IPhi2plus}).
For the function with $n=0$ we need to use the Dirac traces~(\ref{eq:trace1-Phi1},\ref{eq:traceg5-Phi1}) and the Wigner transform~(\ref{eq:op1plus}); for $n=1$ one should apply traces (\ref{eq:trace1-Phi1},\ref{eq:traceg5-Phi1},\ref{eq:trace1-Phi2},\ref{eq:traceg5-Phi2}) and the Wigner transform~(\ref{eq:op2plus}); and for $n=2$ one uses the traces~(\ref{eq:trace1-Phi1},\ref{eq:traceg5-Phi1},\ref{eq:trace1-Phi2},\ref{eq:traceg5-Phi2},\ref{eq:trace1-Phi3},\ref{eq:traceg5-Phi3}) and the Wigner transforms (\ref{eq:op1plus},\ref{eq:op3plus}). In all formulas of the Wigner transform one needs to exchange $\rho_G^\chi$ by $\rho_\Phi^\chi$.

What we find is that for the pieces not depending on chirality (the $\alpha^{(n)}$ operators in Table~\ref{tab:ope}), we get that $\Phi^\chi_{E,v}$ obeys the same form of the equation as $G^\chi_{E,v}$ and the same dispersion law. For the other pieces, the spin operator is replaced with projections over the $n_{\perp,\pm}$ vectors.

%%%%%%%%%%%%%%%%%%%%%%%%%%%%%%%%%%%%%%%
\section{Calculation of fermion scalar density~\label{app:scalarden}}
%%%%%%%%%%%%%%%%%%%%%%%%%%%%%%%%%%%%

In this appendix we provide details on the calculation of the scalar density $\sigma(X)$ in terms of the OSEFT Green functions $G^\chi (X,q)$ and $\Phi^\chi(X,q)$. This procedure contains common steps to the method to derive the kinetic equation and dispersion relation of the different components of the fermion Wigner function. In particular, we always need to calculate some Dirac traces, and perform the Wigner transformation of two-point functions implementing the gradient expansion. 

The starting point is the definition of the scalar density $\sigma(x)$ in terms of the OSEFT fields,
\be
\sigma (x) = \lim_{y \rightarrow x} \textrm{Tr} \left( \langle \bar {\psi} (y) \psi(x) 
\rangle \right) = \lim_{y \rightarrow x} {\rm Tr}\left( \langle [ \bar {\chi}_v (y)+ \bar{H}^{(1)}_{\tilde v} (y) ][   {\chi}_v (x)+ H^{(1)}_{\tilde v} (x) ] \rangle \right) \ , \label{eq:rhodef}
\ee
where the point-splitting regularization allows us to perform the gradient expansion. To the four terms inside the trace one has to include the ${\cal O}(1/E^2)$ term coming from the local field redefinition of the OSEFT field~(\ref{LFR}). Therefore, 
\begin{align} 
\langle \bar {\psi} (y) \psi(x) 
\rangle  & = \langle \bar {\chi}_v (y) \chi_v(x) \rangle + \langle \bar {\chi}_v (y) \chi_v(x) \rangle_{ \rm LFR} + \langle \bar{H}^{(1)}_{\tilde v} (y) {\chi}_v (x) \rangle \nn \\ 
& +\langle \bar {\chi}_v (y) H^{(1)}_{\tilde v} (x) \rangle + 
\langle \bar{H}^{(1)}_{\tilde v} (y)  H^{(1)}_{\tilde v} (x)  \rangle \ . \label{eq:bilinear}
\end{align}
First of all, it is possible to use the property (\ref{eq:noscalar}) to remove the first two terms. A similar use of the projectors in the last term also makes it vanish,
\be \langle \bar{H}^{(1)}_{\tilde v} (y)  H^{(1)}_{\tilde v} (x)  \rangle = \langle \bar{H}^{(1)}_{\tilde v} (y)  P_{v} \  P_{\tilde v} H_{\tilde v}^{(1)}(x) \rangle = 0 \ . \ee
There are only two remaining crossed terms in Eq.~(\ref{eq:bilinear}). Then, one uses Eq.~(\ref{eq:Hfield}) to replace the field $H^{(1)}_{\tilde{v}}(x)$ in terms of $\chi_v(x)$. This brings a $1/E$ expansion which we perform up to ${\mathcal O}(1/E^3)$. We find 4 contributions inside the Dirac trace,
\begin{align} 
\langle \bar {\psi} (y) \psi(x) 
\rangle  & = \left\langle \bar {\chi}_v (y) \frac{i \slashed{D}_{\perp,x} +m }{2E} \frac{\slashed{\tilde{v}}}{2} \chi_v(x) \right\rangle  + \left\langle \bar {\chi}_v (y) \frac{-(i \slashed{D}_{\perp,y})^* +m }{2E} \frac{\slashed{\tilde{v}}}{2} \chi_v(x) \right\rangle \nn \\
& - \left\langle \bar{\chi}_v (y) \frac{i \tilde{v} \cdot D_x}{2E} \frac{i \slashed{D}_{\perp,x} +m }{2E} \frac{\slashed{\tilde{v}}}{2} \chi_v(x) \right\rangle  - \left\langle \bar {\chi}_v (y) \frac{-(i \slashed{D}_{\perp,y})^* +m }{2E} \frac{(i \tilde{v} \cdot D_y)^*}{2E} \frac{\slashed{\tilde{v}}}{2} \chi_v(x) \right\rangle \ ,
\end{align}
where all derivative operators with a complex conjugation act to the left i.e. over the $y$ coordinate. Reordering terms one can write
\begin{align} 
\langle \bar {\psi} (y) \psi(x) \rangle  & = 
\frac{1}{2E} \left\{ \left[ iD^\mu_x - (iD^\mu_y)^* \right] 
\left\langle \bar {\chi}_v (y) \gamma_{\perp,\mu} \frac{\slashed{\tilde{v}}}{2} \chi_v(x) \right\rangle + 2m \left\langle \bar {\chi}_v (y) \frac{\slashed{\tilde{v}}}{2} \chi_v(x) \right\rangle 
\right\} \nn \\
& - \frac{1}{4E^2} \left\{ \tilde{v}_\alpha \left[ iD^\alpha_x iD^\mu_x - (iD^\mu_y)^*(iD^\alpha_y)^*\right] 
\left\langle \bar {\chi}_v (y) \gamma_{\perp,\mu} \frac{\slashed{\tilde{v}}}{2} \chi_v(x) \right\rangle \right. \nn \\
& + \left. \tilde{v}_\alpha m \left[ iD^\alpha_x + (iD^\alpha_y)^* \right] \left\langle \bar {\chi}_v (y) \frac{\slashed{\tilde{v}}}{2} \chi_v(x) \right\rangle \right\} \ . 
\end{align}
Following Eq.~(\ref{eq:rhodef}) we need to take the Dirac trace in both terms. Then, all the 2-point functions can be written in terms of known traces given in App.~\ref{app:details},
\begin{align} 
\textrm{Tr} \langle \bar {\psi} (y) \psi(x) \rangle  & = 
\frac{1}{2E} \left\{ \left[ iD^\mu_x - (iD^\mu_y)^* \right] 
\textrm{Tr} \left[  \gamma_{\perp,\mu} \frac{\slashed{\tilde{v}}}{2} S^<_{E,v} (x,y) \right] + 2m \textrm{Tr} \left[  \frac{\slashed{\tilde{v}}}{2} S^<_{E,v} (x,y) \right] 
\right\} \nn \\
& - \frac{1}{4E^2} \left\{ \tilde{v}_\alpha \left[ iD^\alpha_x iD^\mu_x - (iD^\mu_y)^*(iD^\alpha_y)^*\right] 
\textrm{Tr} \left[ \gamma_{\perp,\mu} \frac{\slashed{\tilde{v}}}{2} S^<_{E,v} (x,y) \right] \right. \nn \\
& + \left. \tilde{v}_\alpha m \left[ iD^\alpha_x + (iD^\alpha_y)^* \right] 
\textrm{Tr} \left[ \frac{\slashed{\tilde{v}}}{2} S^<_{E,v} (x,y) \right] \right\} \nn \\  
&= \frac{1}{2E} \left\{ \left[ iD^\mu_x - (iD^\mu_y)^* \right] 
(-2) \sum_{\chi=\pm} \chi n^\mu_{\perp,\chi} \Phi_{E,v}^\chi (x,y) + 4m \sum_{\chi=\pm} G^\chi_{E,v} (x,y) \right\} \nn \\
& - \frac{1}{4E^2} \left\{ \tilde{v}_\alpha \left[ iD^\alpha_x iD^\mu_x - (iD^\mu_y)^*(iD^\alpha_y)^*\right] 
(-2) \sum_{\chi=\pm} \chi n^\mu_{\perp,\chi} \Phi_{E,v}^\chi (x,y) \right. \nn \\
& + \left. 2 \tilde{v}_\alpha m \left[ iD^\alpha_x + (iD^\alpha_y)^* \right] 
\sum_{\chi=\pm} G^\chi_{E,v} (x,y) \right\} \ . \label{eq:traceofbil}
\end{align}
The traces are now written in terms of the OSEFT functions $G^\chi_{E,v} (x,y)$ and $\Phi_{E,v}^\chi (x,y)$, and some differential operators acting on them.

To continue with the calculation we will make use of the gradient expansion, which is performed together with a Wigner transform of the fields in both sides. Details about the Wigner transform have been reviewed in App.~\ref{app:Wigner}.

We summarize here how the different operators transform under Wigner transformation,
\begin{align}
 \left[ iD^\mu_x - (iD^\mu_y)^* \right] \Phi_{E,v}^\chi (x,y) & \WT i \Delta_k^\mu \Phi_{E,v}^\chi (X,k) \\
 m G^\chi_{E,v} (x,y) & \WT m G_{E,v}^\chi(X,k) \\
 \left[ iD^\alpha_x iD^\mu_x -  (iD^\mu_y)^*(iD^\alpha_y)^*\right] \Phi_{E,v}^\chi (x,y) & \WT i[k^\alpha \Delta_k^\mu + k^\mu \Delta_k^\alpha - e F^{\alpha\mu} (X) ] \Phi_{E,v}^\chi(X,k)\\
 m \left[ iD^\alpha_x + (iD^\alpha_y)^* \right] G^\chi_{E,v} (x,y) & \WT 2mk^\alpha G^\chi_{E,v} (X,k) \ ,
\end{align}
where the final functions depend on the center-of-mass coordinate $X$ and the kinetic (residual) momentum $k$.

We insert these terms into the Wigner transform of Eq.~(\ref{eq:traceofbil}) to obtain
\begin{align} \textrm{Tr} \langle \bar {\psi} \psi \rangle (X,k) 
&= \frac{1}{2E} \left\{ -2 i \sum_{\chi=\pm} \chi n_{\perp,\chi} \cdot \Delta_k \Phi_{E,v}^\chi (X,k) + 4m \sum_{\chi=\pm} G^\chi_{E,v} (X,k) \right\} \nn \\
& + \frac{1}{2E^2} \left\{   i\left[ \tilde{v} \cdot k \Delta_k^\mu + k^\mu \tilde{v} \cdot \Delta_k - e \tilde{v}_\alpha eF^{\alpha\mu} (X) \right] 
 \sum_{\chi=\pm} \chi n_{\perp,\chi,\mu} \Phi_{E,v}^\chi (X,k) \right. \nn \\
& - \left. 2m\tilde{v} \cdot k \sum_{\chi=\pm} G^\chi_{E,v} (X,k) \right\} \ .
\end{align}

The definition of the scalar density is given in terms of this trace,
\be \sigma(X) = \sum_{E,v} \int \frac{d^4k}{(2\pi)^4}  \textrm{Tr} \langle \bar {\psi} \psi \rangle (X,k) \ , \ee
which reads
\begin{align} 
\sigma(X) & = \sum_{\chi=\pm} \sum_{E,v} \int \frac{d^4k}{(2\pi)^4} \frac{1}{E} \left[ 2m  G^\chi_{E,v} (X,k)  - \chi i  (n_{\perp,\chi} \cdot \Delta_k) \Phi_{E,v}^\chi (X,k)  \right] \nn \\
& + \frac{1}{2E^2} \left[ - 2m \tilde{v} \cdot  k  G^\chi_{E,v} (X,k) + \chi i\left( \tilde{v} \cdot k n_{\perp,\chi} \cdot \Delta_k + k \cdot n_{\perp,\chi} \tilde{v} \cdot \Delta_k - e \tilde{v}^\alpha F_{\alpha\mu} (X) n_{\perp,\chi}^\mu \right) \Phi_{E,v}^\chi (X,k)  \right] \ .
\end{align}
This is the expression presented in Eq.~(\ref{eq:sigmaresidual}) in terms of the residual momentum. The result in terms of the full momentum is given in Eq.~(\ref{eq:sigmafull}).

\section{Scalar density under Type II transformation~\label{app:typeIISigma}}

We detail the action of a Type II transformation onto the scalar fermion density $\sigma(X)$ and show that it remains invariant upon integration over the phase space variables.

We start with the expression of the scalar density in terms of the residual momentum in Eq.~(\ref{eq:sigmaresidual}). Many of the terms change under a Type II transformation $\delta_{(II)}$. Keeping terms up to ${\cal O} (1/E^3)$ the different factors to be transformed are
\begin{align}
\delta_{(II)} \sigma(X) & =     \sum_{\chi=\pm} \sum_{E,v} \int \frac{d^4k}{(2 \pi)^4}  \left[ \frac {2m}{E} \delta G^\chi_{E,v}(X,k)-  \chi \frac{i}{E} (\delta n^\chi_\perp \cdot \Delta_k) \Phi_{E,v}^\chi (X,k)  \right. \nn \\
& \left. -  \chi \frac{i}{E} (n^\chi_\perp \cdot \Delta_k) \delta \Phi_{E,v}^\chi (X,k)   -\frac{m}{E^2} \delta \tilde{v} \cdot k G^\chi_{E,v}(X,k)-\frac{m}{E^2} \tilde{v} \cdot k \delta G^\chi_{E,v}(X,k) \right. \nn \\
&+ \left. \chi \frac{i}{2E^2}   \left( \delta \tilde{v} \cdot k \ n_{\perp}^\chi \cdot \Delta+\tilde{v} \cdot k \ \delta n_{\perp}^\chi \cdot \Delta_k + k\cdot \delta n_{\perp}^\chi \tilde{v} \cdot \Delta_k + k\cdot n_{\perp}^\chi \delta\tilde{v} \cdot \Delta_k \right. \right. \nn \\
& - \left. \left.  e\delta \tilde{v}^\alpha F_{\alpha\mu} (X) n_{\perp,\chi}^{\mu} - e\tilde{v}^\alpha F_{\alpha \mu} (X)\delta n^\mu_{\perp,\chi} \right) \Phi_{E,v}^\chi (X,k) \right. \nn \\
&+ \left. \chi \frac{i}{2E^2}   \left( \tilde{v} \cdot k \ n_{\perp}^\chi \cdot \Delta_k + k\cdot n_{\perp}^\chi \ \tilde{v} \cdot \Delta_k - e \tilde{v}^\alpha F_{\alpha\mu}(X) n_{\perp,\chi}^{\mu} \right) \delta \Phi_{E,v}^\chi (X,k) \right] +  {\cal O} \left( \frac {1}{E^3} \right) \ ,
\end{align}
where we used the fact that the operators $\Delta_k^\mu, k^\mu$ and $F_{\alpha \mu}(X)$ do not change under type II transformations. The remaining factors do transform according to the rules given in Ref.~\cite{Carignano:2018gqt}, together with the transformation of $n^\mu_{\perp,\chi}$ given in Eq.~(\ref{eq:typeIInperp}); and the transformation properties of the Wigner functions in Eqs.~(\ref{eq:GvtypeII}) and (\ref{eq:PhitypeII}).

After some initial simplifications one obtains,
\begin{align}
\delta_{(II)} \sigma(X) & =  \sum_{\chi=\pm} \sum_{E,v} \int \frac{d^4k}{(2 \pi)^4}  \left\{ \frac {2m}{E} \left( - \frac{\chi}{8E} \epsilon^{\mu_\perp \nu_\perp \alpha \beta } v_\alpha \tilde{v}_\beta \epsilon^\perp_\nu \Delta_{k,\mu} G_{E,v}^\chi (X,k) \right. \right. \nn \\
&+ \left. \left. \frac{\chi m}{4E} \epsilon_\perp \cdot [n_{\perp,+} \Phi_{E,v}^\dag (X,k)+ n_{\perp,-} \Phi_{E,v} (X,k)] \right) \right. \nn \\ 
& \left. - \chi \frac{\epsilon_\perp \cdot n_{\perp,\chi}}{2} \frac{i}{E^2} k_\perp \cdot \Delta_k \Phi_{E,v}^\chi (X,k) +\chi \frac{ie}{4E^2} \epsilon_\perp \cdot n_{\perp,\chi} \tilde{v}^\alpha F_{\alpha \mu}(X) v^\mu \Phi^\chi_{E,v}(X,k)\right. \nn \\
& \left. -  \chi \frac{i}{E} (n^\chi_\perp \cdot \Delta_k) \left[ - \frac{\chi}{4E} \epsilon^{\mu_\perp \nu_\perp \alpha \beta } v_\alpha \tilde{v}_\beta
\epsilon^\perp_\mu (i k_\nu) 
\Phi_{E,v}^\chi - \frac{m}{4E} \epsilon_\perp \cdot   n_{\perp,-\chi} \ (G^R_{E,v} - G^{L}_{E,v} ) \right]     \right. \nn \\
&+ \left. \chi \frac{i}{2E^2}   \left(  k\cdot n_{\perp}^\chi \epsilon_\perp \cdot \Delta_k -e \epsilon_\perp^\alpha F_{\alpha\mu}(X) n_{\perp,\chi}^{\mu} \right) \Phi_{E,v}^\chi (X,k)  \right\} +  {\cal O} \left( \frac {1}{E^3} \right) \ ,
\end{align}
where we have applied the kinetic equation for $\Phi^\chi_{E,v}$ (\ref{transeqk-Phi}) and its on-shell condition $K_m^\chi=0$.

Notice that the second line in the previous equation vanishes when both chiralities are summed. 

One can perform more simplifications by decomposing the perpendicular vector $k_\perp^\mu,\epsilon_\perp^\mu$ as
\begin{align} 
 k_\perp^\mu = - \frac{k_\perp \cdot n_{\perp,+}}{2}n_{\perp,-}^\mu - \frac{k_\perp \cdot n_{\perp,-}}{2} n_{\perp,+}^\mu \ , \\ 
 \epsilon_\perp^\mu = - \frac{\epsilon_\perp \cdot n_{\perp,+}}{2}n_{\perp,-}^\mu - \frac{\epsilon_\perp \cdot n_{\perp,-}}{2} n_{\perp,+}^\mu \ , 
\end{align}
together with the relation
\be i \epsilon^{\mu_\perp \nu \alpha \beta  } v_\alpha  n^\perp_{\nu,\chi} = \chi v^\beta n_{\perp,\chi}^\mu \ . \ee 

Using this relation it is possible to see how several terms cancel, in particular all those depending on $G_{E,v}^\chi(X,k)$. Only two terms proportional to the electromagnetic tensor remain 
\begin{align}
\delta_{(II)} \sigma(X) & =     \sum_{\chi=\pm} \sum_{E,v} \int \frac{d^4k}{(2 \pi)^4}  
  \frac{\chi i e}{4E^2}   \epsilon_\perp \cdot n_{\perp,\chi} F_{\alpha \mu} (X) (\tilde{v}^\alpha v^\mu + n_{\perp,-\chi}^\alpha n_{\perp,\chi}^{\mu})  \Phi_{E,v}^\chi (X,k)   +  {\cal O} \left( \frac {1}{E^3} \right) \ ,
\end{align}

If we apply the following identities,
\begin{align}
F_{\alpha \mu}(X) \tilde{v}^\alpha v^\mu &= F_{\alpha \mu} (X) (v\wedge u)^{\mu\alpha} \ , \\
F_{\alpha \mu}(X) n^\alpha_{\perp,-\chi} n^\mu_{\perp,\chi} & = \frac{\chi}{2} F_{\alpha \mu}(X) (n_{\perp,+} \wedge n_{\perp,-} )^{\mu\alpha} \ , \\ 
 \frac12 (n_{\perp,+} \wedge n_{\perp,-})^{\mu\alpha} & =- [(v \wedge u)^{\mu \alpha}]^\star 
\end{align}
we can write
\begin{align}
\delta_{(II)} \sigma(X) & =     \sum_{\chi=\pm} \sum_{E,v} \int \frac{d^4k}{(2 \pi)^4}  
  \frac{\chi i}{4E^2}   \epsilon_\perp \cdot n_{\perp,\chi} F_{\alpha \mu}(X) \left\{ (v \wedge u)^{\mu\alpha} -\chi [(v\wedge u)^{\mu\alpha}]^\star \right\}   \Phi_{E,v}^\chi (X,k)   +  {\cal O} \left( \frac {1}{E^3} \right) \ .
\end{align}

In terms of the full momentum the transformation reads,
\be \delta_{(II)}\sigma(X) = \sum_{\chi=\pm} \int \frac{d^4q}{(2\pi)^4}  \frac{\chi i e}{4E_q^2}  \epsilon_\perp \cdot n_{\perp,\chi,q} F_{\alpha \mu} (X)
\left\{ (v_\chi \wedge u)^{\mu \alpha} -\chi [(v_\chi \wedge u)^{\mu\alpha}]^\star \right\} \Phi^\chi (X,q) + {\cal O} \left( \frac{1}{E_q^3} \right) \ . \ee

By symmetry of the integrand when using an isotropic $\Phi_{E,v}^\chi (X,q)$, the integral vanishes and $\sigma(X)$ remains invariant under Type II transformations. As the density is a Lorentz scalar it is not that surprising that in particular, it should be invariant under the Lorentz transformation encoded in the RI Type II transformation (combination of a boost and a rotation as explained in Ref.~\cite{Carignano:2018gqt}). 

\end{appendix}

\end{document}